\documentclass[final, twocolumn, a4paper]{aastex631}

\usepackage{color}
\usepackage{enumitem}
\usepackage{hyperref}
\usepackage{verbatim}
\usepackage{amsmath,amstext,mathrsfs}
\usepackage[all]{hypcap} 
\usepackage{afterpage}    
\usepackage{marginnote}
\usepackage{makecell}
\usepackage{subfigure}
\usepackage{multirow}

\shorttitle{Magnetic field in giant filaments}
    \shortauthors{Zhao et al}
\usepackage{enumitem}
\setlist[enumerate]{listparindent=\parindent}
\makeatletter

\newcommand{\Rmnum}[1]{\expandafter\slowromancap\romannumeral #1@}
\makeatother

\begin{document}
    \hfuzz = 150pt
\title{\Large\bfseries Magnetic Fields in Giant Filaments Probed by the Velocity Gradient Technique:  Regular Magnetic Field interrupted by Magnetization Gaps}

\correspondingauthor{Mengke Zhao,Guang-Xing Li, Jianjun Zhou}
\email{mkzhao628@gmail.com, gxli@ynu.edu.cn, zhoujj@xao.ac.cn}

\author[0000-0003-0596-6608]{Mengke Zhao}
\affil{Xinjiang Astronomical Observatory, Chinese Academy of Sciences,  Urumqi, 830011, People's Republic of China}
\affil{University of Chinese Academy of Sciences, Beijing, 100049, People's Republic of China}

\author[0000-0003-3144-1952]{Guang-Xing Li}
\affil{South-Western Institute for Astronomy Research, Yunnan University, Kunming, 650500 Yunnan, People's Republic of China}

\author[0000-0003-0356-818X]{Jianjun Zhou}
\affil{Xinjiang Astronomical Observatory, Chinese Academy of Sciences,  Urumqi, 830011, People's Republic of China}
\affil{Key Laboratory of Radio Astronomy, Chinese Academy of Sciences  Urumqi,830011, People's Republic of China}
\affil{Xinjiang Key Laboratory of Radio Astrophysics, Urumqi 830011, People's Republic of China}

\author[0000-0002-4154-4309]{Xindi Tang}
\affil{Xinjiang Astronomical Observatory, Chinese Academy of Sciences,  Urumqi, 830011, People's Republic of China}
\affil{University of Chinese Academy of Sciences, Beijing, 100049, People's Republic of China}
\affil{Key Laboratory of Radio Astronomy, Chinese Academy of Sciences  Urumqi,830011, People's Republic of China}
\affil{Xinjiang Key Laboratory of Radio Astrophysics, Urumqi 830011, People's Republic of China}

\author{Jarken Esimbek}
\affil{Xinjiang Astronomical Observatory, Chinese Academy of Sciences,  Urumqi, 830011, People's Republic of China}
\affil{Key Laboratory of Radio Astronomy, Chinese Academy of Sciences  Urumqi,830011, People's Republic of China}
\affil{Xinjiang Key Laboratory of Radio Astrophysics, Urumqi 830011, People's Republic of China}
\author[0000-0002-8760-8988]{Yuxin He}
\affil{Xinjiang Astronomical Observatory, Chinese Academy of Sciences,  Urumqi, 830011, People's Republic of China}
\affil{University of Chinese Academy of Sciences, Beijing, 100049, People's Republic of China}
\affil{Key Laboratory of Radio Astronomy, Chinese Academy of Sciences  Urumqi,830011, People's Republic of China}
\affil{Xinjiang Key Laboratory of Radio Astrophysics, Urumqi 830011, People's Republic of China}
\author{Dalei Li}
\affil{Xinjiang Astronomical Observatory, Chinese Academy of Sciences,  Urumqi, 830011, People's Republic of China}
\affil{University of Chinese Academy of Sciences, Beijing, 100049, People's Republic of China}
\affil{Key Laboratory of Radio Astronomy, Chinese Academy of Sciences  Urumqi,830011, People's Republic of China}
\affil{Xinjiang Key Laboratory of Radio Astrophysics, Urumqi 830011, People's Republic of China}
\author{Weiguang Ji}
\affil{Xinjiang Astronomical Observatory, Chinese Academy of Sciences,  Urumqi, 830011, People's Republic of China}]
\author{Zhengxue Chang}
\affil{College of Mathematics and Physics, Handan University, No.530 Xueyuan Road, Hanshang District, 056005 Handan, PR China}
\author{Kadirya Tursun}
\affil{Xinjiang Astronomical Observatory, Chinese Academy of Sciences,  Urumqi, 830011, People's Republic of China}

\begin{abstract}
We study the magnetic field structures in six giant filaments associated with the spiral arms of the Milky Way by applying the Velocity Gradient technique (VGT) to the $^{13}$CO spectroscopic data from GRS, Fugin, and SEDIGSM surveys.
Compared to dust polarized emission, the VGT allows us to separate the foreground and background using the velocity information, from which the orientation of the magnetic field can be reliably determined. 
We find that in the most cases, the magnetic fields stay aligned with the filament bodies, which are parallel to the disk midplane. 
Among these, G29, G47, and G51 exhibit smooth magnetic fields, and G24, G339, and G349 exhibit discontinuities.
The fact that most filaments have magnetic fields that stay aligned with the Galactic disk midplane suggests that Galactic shear can be responsible for shaping the filaments. 
The fact that the magnetic field can stay regular at the resolution of our analysis ($\lesssim$  10 pc) where the turbulence crossing time is short compared to the shear time suggests that turbulent motion can not effectively disrupt the regular orientation of the magnetic field. 
The discontinuities found in some filaments can be caused by processes including filament reassembly, gravitational collapse, and stellar feedback.
\end{abstract}

\section{Introduction}
Observationally, giant molecular clouds are the densest and coldest phase in the Milky Way \citep{1969ApJ...155L.149F,1977ApJ...218..148M}. 
They have complex, filamentary morphology \citep{1979ApJS...41...87S,1987ApJ...312L..45B,2000prpl.conf...97W,2008A&A...487..993K,2010A&A...518L.103M,2018ApJ...864..153Z}.
On the Galactic scale, molecular clouds are organized in Galactic-sized giant filaments whose lengths can exceed the scale height of the Galactic disk \citep{2013A&A...559A..34L}. 
These objects contain information on the dynamical interplay between the Galactic disks and the clouds \citep{2022MNRAS.tmpL..71L}.

Recent studies suggest that the magnetic field could play a key role in the
collapse of molecular clouds and star formation
\citep{2012ARA&A..50...29C,2019FrASS...6....3H,2021Galax...9...41L}. Despite
significant processes, the importance of the magnetic field on the giant molecular filaments remains unclear.
Measuring the magnetic field is still a challenging task. This is because continuum mission from dust,  an
effective tracer for magnetic field orientation, is no longer reliable, due to
the presence of foreground and background emission in the disk mid-plane.

Recently, a new method called the Velocity Gradient Technique (VGT; \citealt{2017ApJ...835...41G, 2018ApJ...853...96L,2023MNRAS.tmp.1894H}) has recently emerged, offering a fresh approach to estimating magnetic field orientation by analyzing velocity-resolved maps that trace the gas.
The VGT is based on the theory of the anisotropy of magneto-hydrodynamic (MHD) turbulence, where turbulent eddies tend to stretch along the magnetic field lines \citep{1995ApJ...438..763G, 1999ApJ...517..700L}.
In a magnetized and turbulent medium, these eddies are expected to align with the local magnetic field, and the velocity gradient is perpendicular to the major axis of the eddies \citep{2000ApJ...539..273C, 2001ApJ...554.1175M, 2002ApJ...564..291C}.
By examining the alignment between the magnetic field orientation and the velocity gradient of the turbulence eddies, it becomes possible to infer the magnetic field through the motion of these eddies.
The validity of this approach has been demonstrated through MHD simulations and observations of turbulent media \citep{2019NatAs...3..776H, 2019ApJ...886...17H, 2021MNRAS.502.1768H, 2021ApJ...912....2H, 2022ApJ...934...45Z,2023MNRAS.tmp.1894H}.
Taking advantage of this technique, we measure the magnetic fields of six giant filaments in the Milky Way and study the relation between evolution and the magnetic fields in these giant filaments. 

\begin{figure*}
    \centering
    
        \includegraphics[width = 14cm]{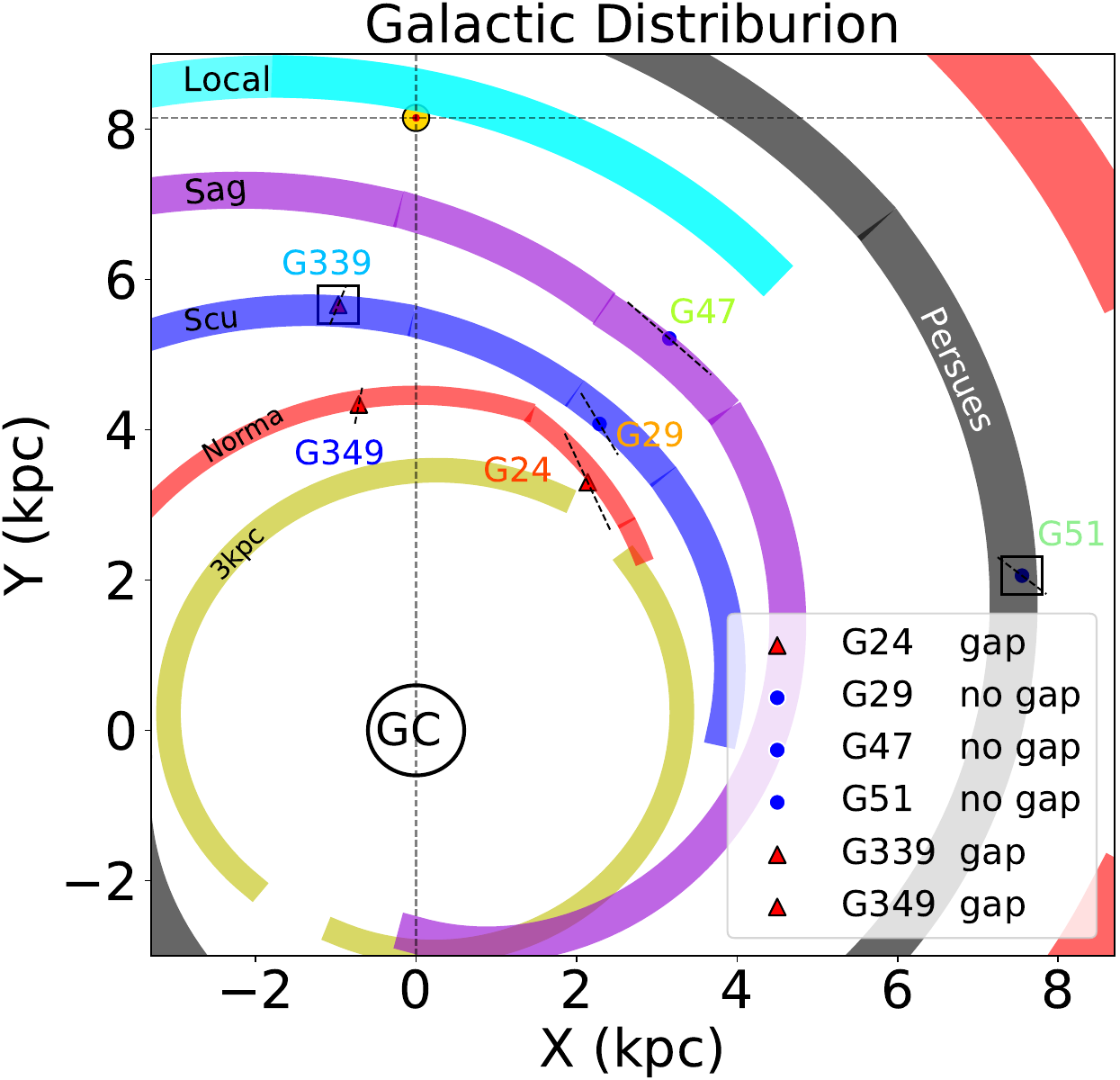}

        \includegraphics[width= 16cm]{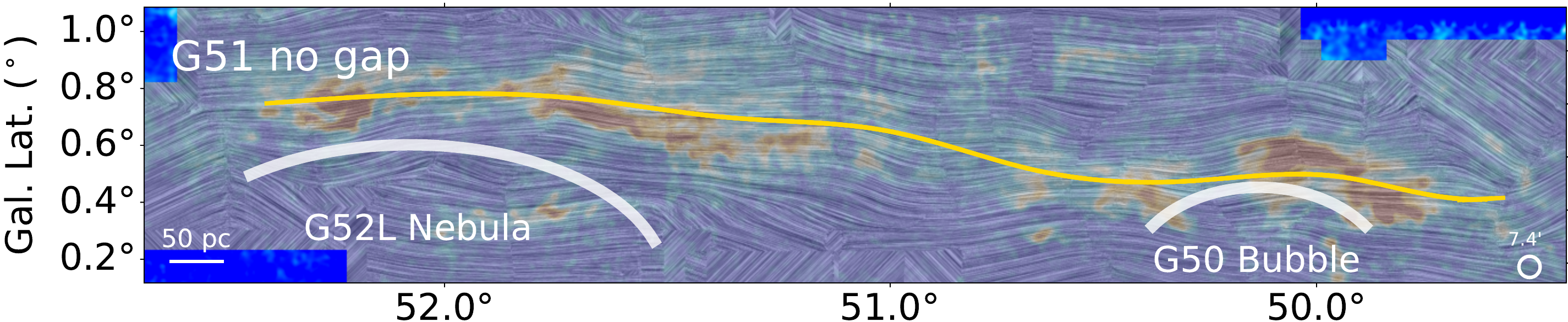}
        \includegraphics[width= 16cm]{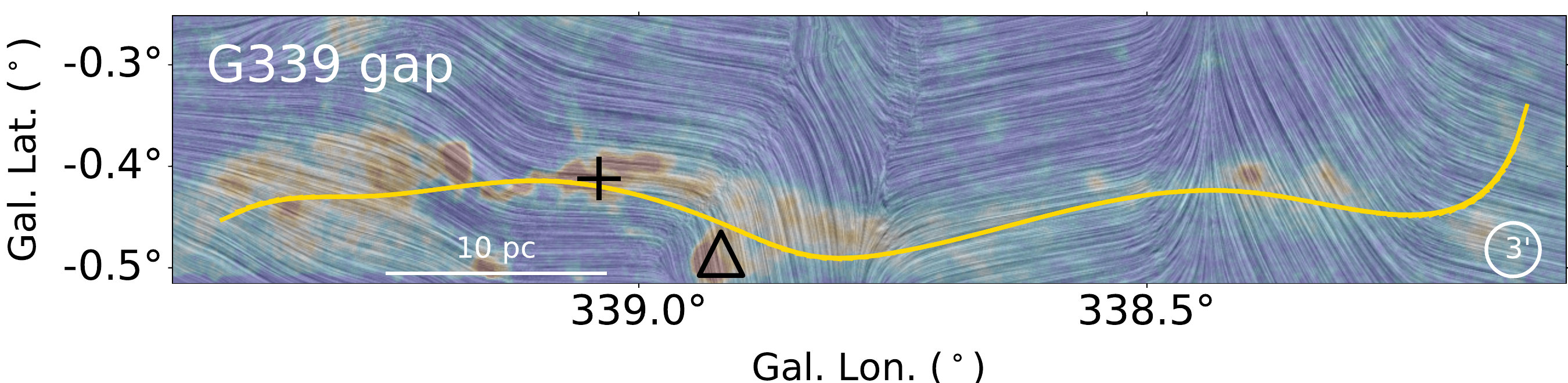}
    \caption{
    \textbf{Types and Galactic distribution  of six large scale
    filaments}. 
    The top panel shows the location of six giant filaments in the galaxy Milky Way\citep{2019ApJ...885..131R}, where the locations of spiral arms are also indicated. 
    The markers with different colors show the position of the filament in the Milky Way. The black dashed lines are error bars of filament distance.
    The triangles indicate the locations of the magnetization gaps.
    The bottom two maps show magnetic field structure from two different types of filaments with and without gaps, which the map of magnetic field orientations of filaments are indicated as line-integral-convolution \citep[LIC]{Cabral_lic} map. 
    The background displays the intensity map from the CO emissions of filaments.
    }
    \label{GF}
\end{figure*}

\begin{table*}
    \centering
    \caption{Physical parameters of the six giant filaments.}
    \begin{tabular}{c  c  c  c  c  c  c  c  c }
        \hline
	    \hline
         Property & Unit & G24 & G29 & G47 & G51 & G339 & G349  \\
         \hline 
         Position($l,b$) & {\rm degree}, {\rm degree} &  024.00,+0.48 & 029.18,$-$0.34 & 047.06,+0.26 & 051.25,+0.6 & 338.47,$-$0.47& 349.4,+0.10  \\
         Velocity Range & km s$^{-1}$ & 96.0$\pm$3.0 & 93.5 $\pm$ 4.5 & 57.5$\pm$3.5 & 4.0$\pm$10.0& -38.0$\pm$3.0 & -63.0$\pm$4.0 \\
         Distance (d) & kpc & 5.3$\pm$0.7 & 4.7 $\pm$ 0.5 & 4.3$\pm$ 0.7 & 9.7 $\pm$ 0.4 & 2.7 $\pm$ 0.3 & 3.88 $\pm$ 0.26 \\
         Arm &   & Norma & Sct-Cen & Sgr-Car & Persus & Sct-Cen & Norma \\
         Galactocentric distance & kpc & 3.93 & 4.67 & 6.09 & 7.83 & 5.74 & 4.39 \\
         Length & pc & 82 & 42 & 62 & 621 & 80 & 161 \\
         Thickness & pc & 4.28 & 2.84 & 2.64 & 21.64  & 3.41 & 3.55 \\
         Velocity Dispersion &  km s$^{-1}$ & 1.53 & 2.33& 1.65& 3.96& 1.45& 2.65 \\
         Galactic Altitude & pc & 44 & 28 & 20 & 102 & 22 & 7 \\
         $\theta_{\rm GP}$  &  {\rm degree} & 0.36 & -0.46 & 0.14 & 0.48 & -0.61 & -0.02  \\
         Mean N$\rm _{H_2}$ & 10$^{22}$ cm$^{-2}$ & 1.09 & 0.90 & 0.70 & 0.41 & 0.97 & 1.16 \\
         Max N$\rm _{H_2}$ & 10$^{22}$ cm$^{-2}$ & 1.41 & 1.18 & 1.07 & 0.71 & 1.39 & 8.10 \\
         Mass & 10$^4$ M$_{\sun}$ & 4.3 & 1.2 & 1.3 & 11.2 & 3.0& 7.4\\
         B-field Beam &   & 2.83$'$ & 2.83$'$ & 2.83$'$ & 7.4$'$ & 3.17$'$ & 3.17$'$ \\
         Resolution & pc & 4.4 & 3.9 & 3.6 & 20.9 & 2.5 & 3.6 \\
         p$_{\rm para}$ &   & 87\% & 100\% & 73 \% & 100\% & 63\% & 51\% \\
         Cross Time & Myr & 52.44 & 17.64 & 32.62 & 153.44 & 53.98 & 59.45 \\
         B-field Cross Time & Myr & 2.81 & 1.64 & 2.13 & 5.16 & 1.69 & 1.33 \\
         \hline
         \hline
    \end{tabular}
    \medskip
    
    \it{Notes:} \rm The position and velocity range of our sample are taken from existing catalogs, \citealt{2013A&A...559A..34L,2015MNRAS.450.4043W,2018ApJ...864..152Z}. 
    Distance and Arms are calculated by the model from Bessel \citep{2016ApJ...823...77R,2019ApJ...885..131R} using the position and velocity range. 
    Using the above parameters, the Galactocentric distance, length, thickness, and Galactic altitude will be derived.
   The calculation of $\theta_{GP}$, which represents the angle between the Galactic Plane and the object, involves subtracting 0.12$^\circ$ from the value of b (see Fig. \ref{figGP}).
    The mean N$\rm _{H_2}$, max N$\rm _{H_2}$, and mass are calculated by the H$_2$ column density, which is fitting by the Spectral Energy Distributed (SED) model from Herschel observations \citep{2010A&A...518L...2P,2010A&A...518L...3G} (detail in Sect.\,\ref{B.2}).
    The B-field beam is the spatial resolution of the magnetic field measured with the VGT method.
    The resolution of the magnetic field is the spatial resolution of the magnetic field measured with VGT.
    The computing process of two cross time is shown in Sect.\,\ref{sec5.2},\ref{sec5.3}. 
    \label{tab1}
\end{table*}

 \section{Achieved Data}\label{data}


\subsection{Spectral Line}

We use the $^{13}$CO rotational transitions to trace the magnetic field by VGT, which $^{13}$CO emission is a good tracer of the dynamical processes and bulk molecular gas at density $\textgreater$ 100 cm$^{-3}$.
Our spectral data comes from three Surveys: Galactic Ring Survey (GRS; $^{13}$CO\,(1-0),
observed by FCRAO, beam size $\sim$ 46$''$
; \citealt{2006ApJS..163..145J}), 
FOREST Unbiased Galactic plane Imaging survey (FUGIN; $^{13}$CO\,(1-0), observed by Nobeyama telescope, beam size $\sim$ 20$''$; \citealt{2017PASJ...69...78U}), 
and Structure Excitation and Dynamics of the Inner Galactic Interstellar Medium, (SEDIGISM; $^{13}$CO\,(2-1), observed by APEX, beam size $\sim$ 30$''$; \citealt{2017A&A...601A.124S,2021MNRAS.500.3064S}).

The GRS Survey employs 14 m telescopes at the five College Radio Astronomy Observatory (FCRAO) to observe the spectral line of $^{13}$CO (1-0).The survey utilizes an on-the-fly scanning mode, with each scanning area denoted as $l \times b$ = 6$'$ $\times$ 18$'$ \citep{2006ApJS..163..145J}.
For the Fugin Survey, observations are carried out using the Nobeyama 45-m telescope to capture the spectral lines of $^{12}$CO (1-0), $^{13}$CO (1-0), and C$^{18}$O (1-0). 
The survey divides its observations into 1$^\circ$$\times$1$^\circ$ fields, employing two sets of on-the-fly scanning modes to cover the areas effectively.
To account for scanning effects, scans are performed perpendicular to the Galactic plane \citep{2017PASJ...69...78U}.
The SEDIGISM Survey is executed using the APEX (Atacama Pathfinder Experiment) telescope, which has a diameter of 12 m \citep{2006A&A...454L..13G}. The survey area is divided into 0.5$^\circ$$\times$0.5$^\circ$ fields. 
Each field is double-covered through on-the-fly mapping, scanning along both the Galactic longitude and latitude directions \citep{2017A&A...601A.124S}.

\subsection{Continuum}
Continuum observations of dust are used to derive the filament mass. 
We derive the column density by SED fitting continuum emission from multiple bands (details concerning the SED fitting can be found in Sect.\,\ref{B.2}). 
The Herschel far-IR continuous spectrum covers the entire Galactic plane with $\mid$ b $\mid$  $\le  1 ^\circ$ \citep{2010PASP..122..314M}.
Continuum at 70 ($\pm$ 15) and 160 ($\pm$ 35)\,$\mu$m are obtained from Photodetector Array Camera and Spectrometer (PACS; \cite{2010A&A...518L...2P}) observations and 250, 350 and 500\,$\mu$m data are obtained from Spectral and Photometric Imaging Receiver (SPIRE; \cite{2010A&A...518L...3G} observations, which have two overlapping bands covering 191$\sim$671 $\mu$m. 870$\mu$m data (wavelength $\sim$ 800-950 $\mu$m) used in this paper is taken from the APEX Telescope Large Area Survey of the Galaxy (ATLASGAL; \cite{2009A&A...504..415S}) survey.



\begin{figure*}
    \centering
    \includegraphics[height=4.9cm]{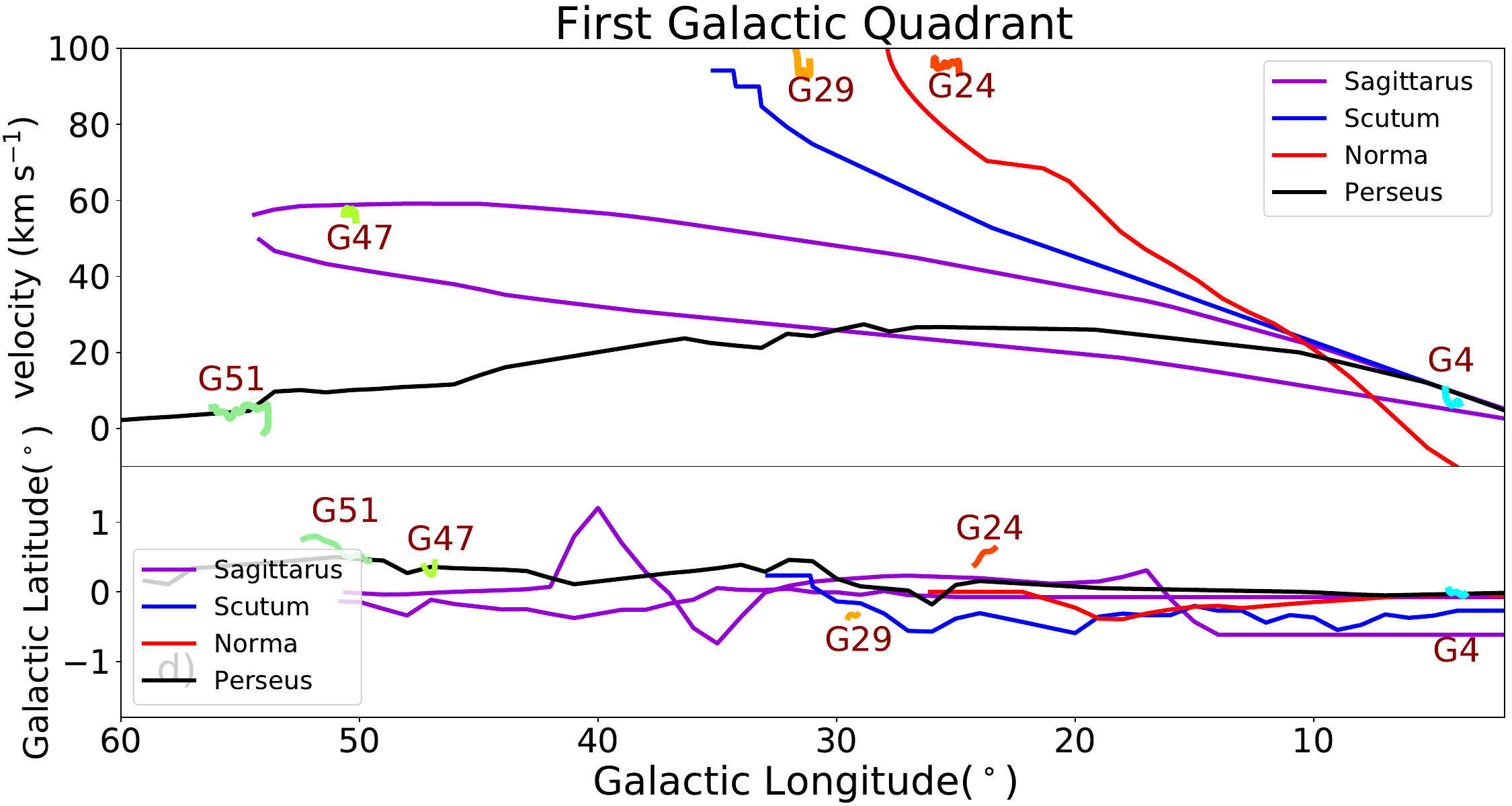}
    \includegraphics[height=4.9cm]{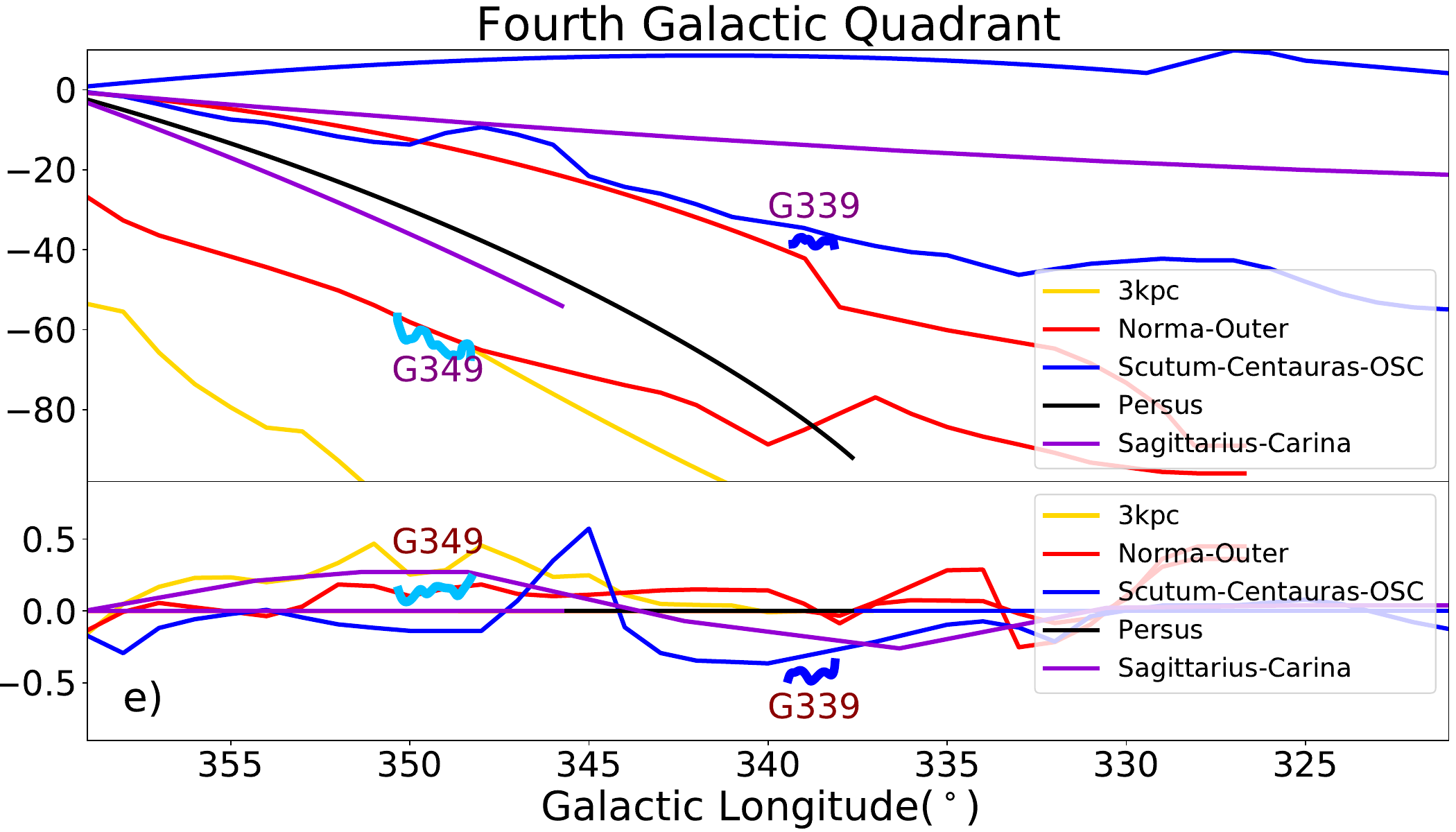}
    \caption{\textbf{$l$-$b$ and $l$-$v$ diagram of spiral arm model where the locations of the giant filaments are indicated}. 
    These panels show distributions of filaments at $l$-$b$ and $l$-$v$ plane at the first Galactic quadrant (right) and fourth Galactic quadrant (left). 
    The long lines with different colors in the label show the positions of spiral arms \citep{2016ApJ...823...77R,2019ApJ...885..131R}.
    The other short lines represent the position of the giant filament of which color is the same as Fig.\,\ref{GF}.
    }
    \label{Galad}
\end{figure*}

\section{method}\label{method}
\subsection{Velocity Gradient Technique}\label{B.1}

In this work, we apply VGT to the $^{13}$CO emission to estimate the magnetic field orientation.

The velocity gradient technique (VGT  \citealt{2017ApJ...835...41G,2018ApJ...853...96L,2018MNRAS.480.1333H}) is the main analysis tool used in this work, which is developed based on the anisotropy of magneto-hydrodynamic turbulence \citep{1995ApJ...438..763G} and fast turbulent reconnection theories \citep{1999ApJ...517..700L}. 
For extracting the velocity information from the position-position-velocity (PPV) cube of the spectral line, the thin velocity channels Ch(x,y) were employed, in which the width of the thin velocity channel is less than the velocity dispersion of the spectral lines observed at the individual line of sight \citep{2023MNRAS.tmp.1894H}. 
We use velocity-resolved maps to study the structure of the magnetic field, using equations Eq.\ref{eq1}, \ref{eq2}, \ref{eq3}, where the velocity-resolved map is the position-position-velocity (PPV) cube of the spectral line, and the results are valid as long as the velocity channel is smaller than the velocity dispersions along the individual line of sights (the cloud is well-resolved in velocity).

\begin{equation}\label{eq1}
    \begin{aligned}
        \rm\bigtriangledown_x Ch_i(x,y) =\frac{Ch_i(x+\Delta x,y) -Ch_i(x-\Delta x,y) }{2\,\Delta x} \,,
    \end{aligned}
\end{equation}
\begin{equation}\label{eq2}
    \begin{aligned}
        \rm\bigtriangledown_y Ch_i(x,y) =\frac{Ch_i(x,y+\Delta y) -Ch_i(x,y-\Delta y) }{2\,\Delta y} \,,
    \end{aligned}
\end{equation}
\begin{equation}\label{eq3}
    \begin{aligned}
        \rm\psi^i_g = tan^{-1} (\frac{\bigtriangledown_y Ch_i(x,y)}{\bigtriangledown_x Ch_i(x,y)}) \,,
    \end{aligned}
\end{equation}
where $\bigtriangledown_x$Ch$_i$(x,y) and $\bigtriangledown_y$Ch$_i$(x,y) are the x and y components of gradient respectively, the $\Delta$x and $\Delta$ y are the x and y components of pixel size of PPV cube. 
The (x,y) display the position in the two-dimensional plane, such as the position (l,b) in this work.
$\psi^i_g$ is raw velocity gradient orientation. 
This is done for pixels whose spectral line emission has a signal-to-noise ratio greater than 3. 
We focus on regions with signal-to-noise ratios larger than 3 to ensure robustness.

For estimating the velocity gradient,a method, sub-block averaging \citep{2017ApJ...837L..24Y}, has been used to export velocity gradients from raw gradients ( intensity gradient in the thin velocity channel) within a sub-block of interest and then plots the corresponding histogram of the raw velocity gradient orientations $\psi^i_g$.
The size of the sub-block is set as 20\,$\times$\,20 pixels, and determines the size of the final magnetic field resolution.
Using the sub-block averaging, the eigen-gradient maps $\psi^i_{g_s}(x,y)$ are taken to calculated the pseudo-Stokes-parameters $Q_g$ and $U_g$.
Then, pseudo-Stokes-parameters $Q_g$ and $U_g$ of the gradient-inferred magnetic field would be constructed by:

\begin{equation}
    \begin{aligned}
        \rm Q_g(x,y) = \sum\limits_{i=1}^{n_v} I_i(x,y)cos(2\psi^i_{g_s}(x,y))  ,
    \end{aligned}
\end{equation}
\begin{equation}
    \begin{aligned}
        \rm U_g(x,y) = \sum\limits_{i=1}^{n_v} I_i(x,y)sin(2\psi^i_{g_s}(x,y)) ,
    \end{aligned}
\end{equation}
\begin{equation}
    \begin{aligned}
        \rm\psi_g = \frac{1}{2}tan^{-1}\frac{U_g}{Q_g} ,
    \end{aligned}
\end{equation}
where I$_i$(x,y) is the two-dimensions (l-b plane) integrated intensity of spectra cubes and n$_v$ is the number of velocity channels. 
The $\psi_g$ is pseudo polarization derived by the pseudo-Stokes-parameters $Q_g$ and $U_g$, which could project three-dimensional data $\psi^i_{g_s}$ into two-dimensions pseudo polarization angle $\psi_g$.
The VGT-derived magnetic field orientation $\psi_B$ is perpendicular to the pseudo polarization angle $\psi_g$ on POS: $\psi_B = \psi_g + \pi /2$ .
This VGT-derived magnetic field orientation is the magnetic field orientation measured with the VGT method by the spectral lines.



\subsection{$\rm H_2$ column density}\label{B.2}

Herschel telescope provides continuum data at 70, 160, 250, 350, and 500$\mu$m continuums.
870$\mu$m continuum is taken from ATLASGAL.
Using the continuums at six bands, the column density distribution is derived by the Spectral Energy Distributed (SED) fitting model. 
Continuums of 70, 160, 250, 350, 500, and 870$\mu$m have been convolved to the same circular Gaussian beam: 35$^\circ$, which is the spatial resolution of 500$\mu$m continuums and the maximum size beam of six band continuum, then use the SED fitting procedure \citep{2015MNRAS.450.4043W} to get column density for each pixel. The intensity at various wavelengths is as:

\begin{equation}
    \begin{aligned}
         I_v = B_v(1 - e^{-\tau_v}) ,
    \end{aligned}
\end{equation}
\begin{equation}
    \begin{aligned}
    \tau_v = \mu_{\rm  H_2} \rm m_{\rm H} \kappa_vN_{\rm H_2} / R_{g_d} ,
    \end{aligned}
\end{equation}
\begin{equation}
    \begin{aligned}
        \rm \kappa_v = 4.0\,(\frac{\emph{v}}{505GHz})^\beta cm^2 g^{-1} ,
    \end{aligned}
\end{equation}
where $B_v$ is Planck function \citep{2008A&A...487..993K}. 
$R_{g_d}$ is the gas to dust ratio around 100. 
Dust emissivity index $\beta$ is 1.75. 
I$_v$ is the flux of continuums.
$\mu_{\rm  H_2}$ is the mean molecular weight for molecular gas.
m$_{\rm H}$ is the mass of the hydrogen atom \citep{2016ApJ...819...66D}.
Here dust temperature and column density are the free parameters. 
In this work, the column density obtained by fitting multi-wavelength continuums is expressed in the number of molecular hydrogen per unit area.
Due to the effect of foreground in the continuum, this column density and the mass derived by column density is the upper limiting of our sample (see Tab.\,\ref{tab1}).

\subsection{Sample}


Our sample included six filaments with relatively simple morphology, such that they serve as probes to physical processes on the Galactic scale such as shear. The sources are selected from previous studies.
These filaments are located at the Galactic disk and reach up to the Galactic scale, whose length exceeds the FWHM thickness of the Galactic disc \citep{2013A&A...559A..34L}.
Among them, five are taken from existing catalogs: G24, G29, and G47 come from \citealt{2015MNRAS.450.4043W}. G339 come from \citealt{2014ApJ...797...53G,2015ApJ...815...23Z}. 
The longest filament G51 has been identified by \citealt{2013A&A...559A..34L}. 
Another filament G349 was found in SEDIGISM Survey \citep{2021MNRAS.500.3064S}. 
The details of our sample, including position and velocity range, are summarized in Table.\,\ref{tab1}. 

The filaments exhibit simple structures, resembling shapes like an "L" or a "C" (detail described in \citealt{2015MNRAS.450.4043W}).
The complex structures may originate from local physical processes at smaller scales, rather than being contributed by the physical processes at galactic scale.
The uncomplicated structures observed in our sample may be attributed to galactic factors.

We study the magnetic field structure of these giant molecular filaments proved by the VGT and explore the relation between the morphology of the magnetic field in plane-of-sky (POS) and other properties such as filament location. 
The VGT technique \citep{2017ApJ...835...41G,2018ApJ...853...96L,2018MNRAS.480.1333H} is a general method to estimate the magnetic field orientation towards molecular gas where MHD turbulence prevails. This is true for filaments selected in this work.

\subsection{Nearest Spiral Arms and Distance}

\begin{figure*}
    \centering
    \includegraphics[height = 6cm]{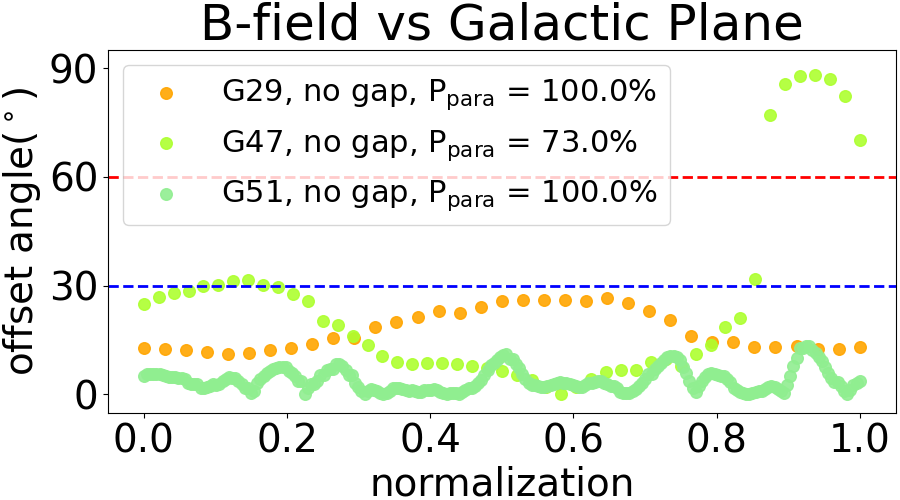}
    \includegraphics[height = 4cm]{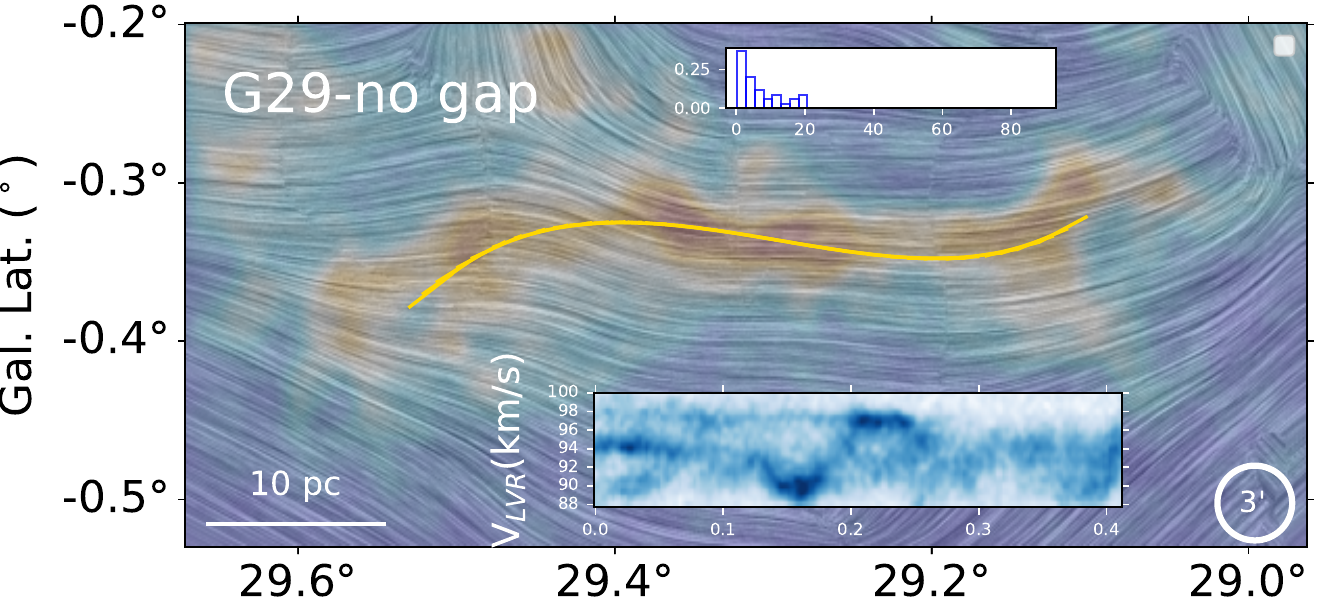}
    \includegraphics[height = 4cm]{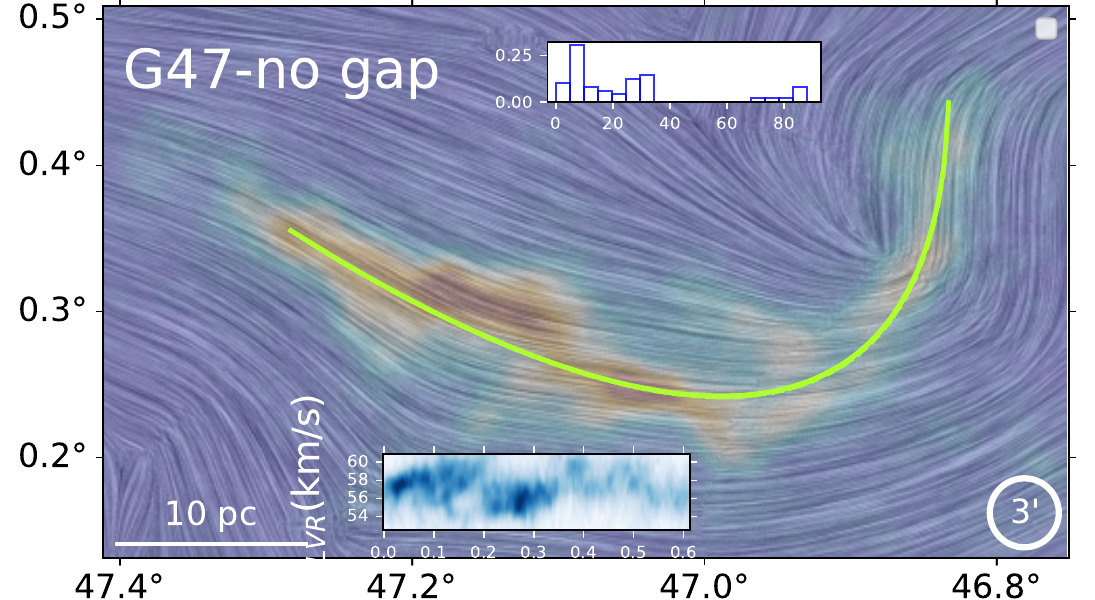}
    \includegraphics[width = 15.2cm]{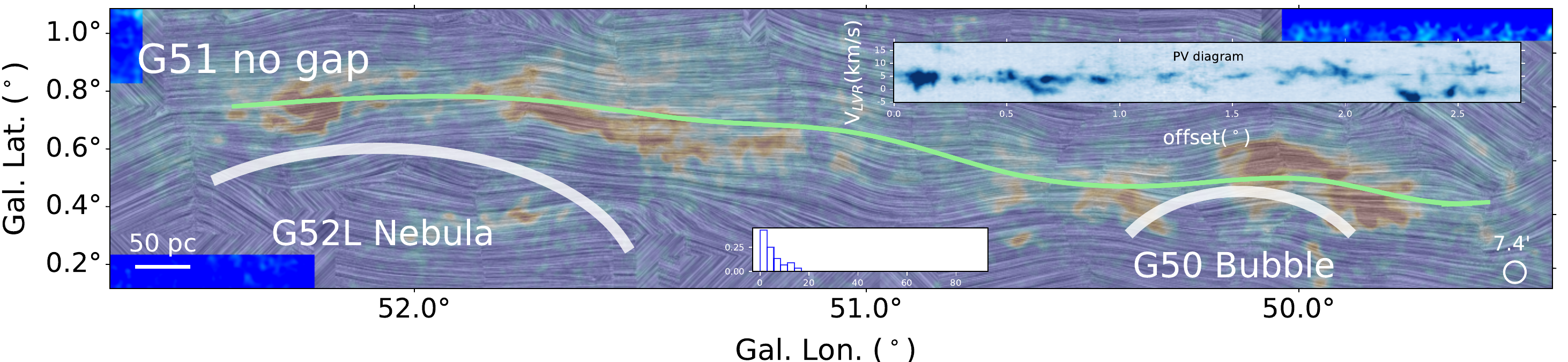}
    \caption{\textbf{Giant filaments with no magnetization gap}. 
    The top panel shows offset angles, between the magnetic field orientation and Galactic Plane, distribution of filaments G29. G47, and G51, which have no gap. 
    The parallel probability p$_{\rm para}$ presents the parallel alignment between B-field and Galactic Plane (see Sect.\ref{BGP}, detail shown in Tab.\,\ref{tab1}). 
    The $x$-axis is the normalized length and the $y$-axis is the absolute offset angle. 
    The middle and bottom panels show the B-field orientations of filaments G29. G47, and G51 by line-integral-convolution (LIC, \citealt{Cabral_lic}). 
    The lines with different colors show the skeleton of filaments whose color is the same as that in Fig.\,\ref{GF}.
    The sub-figure shows the P-V diagram along each filament skeleton.
    The x-axis is filament length unit as degree and the y-axis is velocity V$\rm_{LSR}$ unit as km s$^{-1}$.
    Inset plots represent the normalized distribution of the offset angle between the magnetic field orientation and the Galactic Plane.
    The x-axis is the offset angle ranging from 0$^\circ$ to 90$^\circ$.}
    \label{para}
\end{figure*}

\begin{figure*}
    \centering
    \includegraphics[height = 4.2cm]{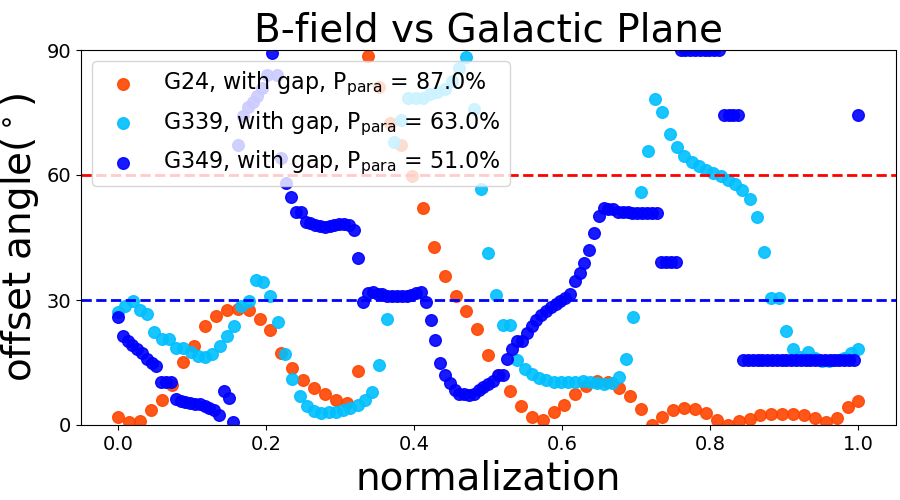}
    \includegraphics[height = 4.2cm]{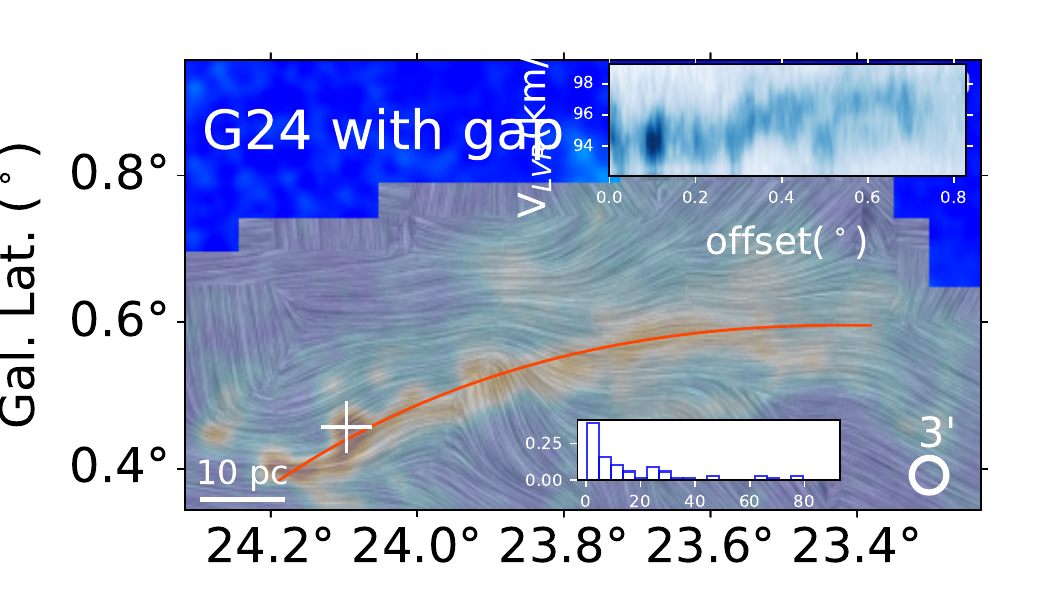}
    \includegraphics[width = 16cm]{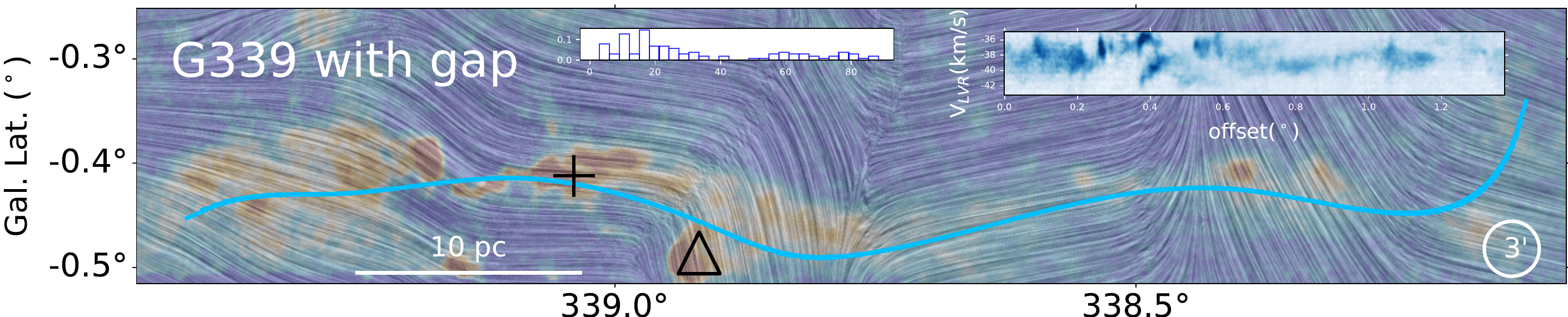}
    \includegraphics[width = 16.6cm]{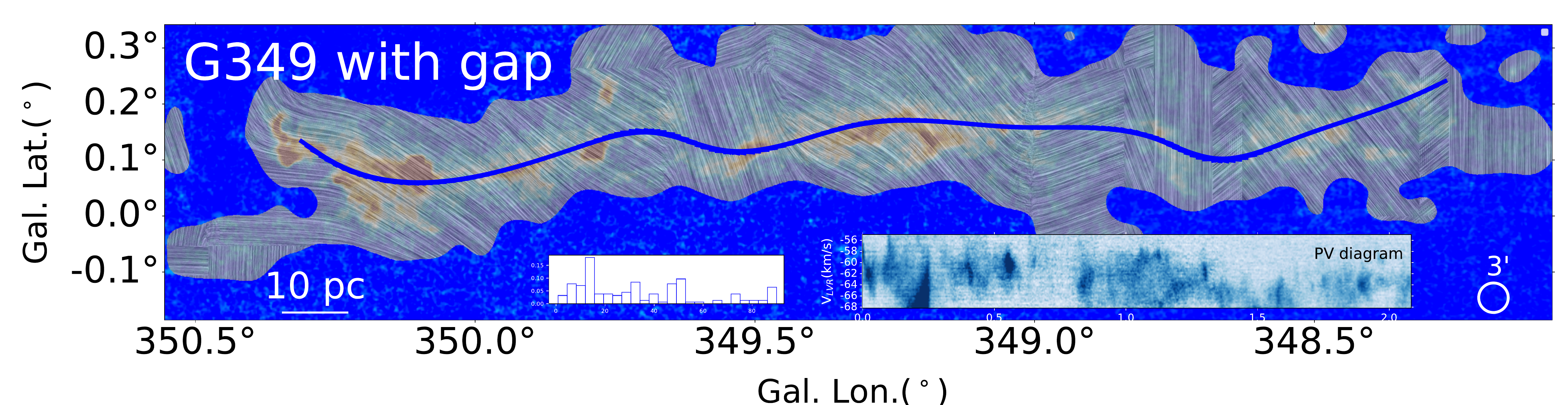}
    \caption{\textbf{Giant filaments with magnetization gaps}. 
    Same as Fig.\,\ref{para} but for filaments, G24, G339, and G349, which have magnetization gaps.
    The cross and circle markers in G24 and G349 display the position of dense clumps identified by \citealt{2014MNRAS.443.1555U}.}
    
    \label{pendi}
\end{figure*}

The positions in the Milky Way of these filaments are shown in Fig.\,\ref{GF}.
The kinematic distance, d, is derived using a Bayesian distance calculator using data from the spiral structure of the Bar and Spiral Structure Legacy (BeSSel) Survey\,\citep{2016ApJ...823...77R,2019ApJ...885..131R}.
The spiral arm models are extracted by the Bayesian distance calculator \citep{2019ApJ...885..131R}. 
To better determine the location of these filaments, the $l-b$ and $l-v$ diagram derived from BeSSeL has been shown in Fig.\,\ref{Galad}. 
According to the illustrations in Fig.\,\ref{GF} and \ref{Galad}, it appears that giant filaments G24 and G29 are potentially positioned within the inter-arm region of the Milky Way and in proximity to nearby spiral arms in the X-Y plane. On the other hand, the remaining four filaments, namely G47, G51, G339, and G349, are situated within the spiral arms themselves.
We use the Python package: Radfil \citep{2018ApJ...864..152Z} to extract the skeleton of the filament body and plot it in the $l-b$ diagram. 
Using the filament's skeleton as input to extract the PV-diagram from the spectral cube using Python package: {\it pvextractor}, the position-velocity (P-V) diagrams are shown in Fig.\,\ref{para}\, and\,\ref{pendi}.  
The length of the filament body is taken as the length of the skeleton. 
Due to the projection effect, the actual filament length can be larger. 
The details including the spiral arms and distance of six giant filaments are illustrated in Table.\,\ref{tab1}.

\section{Result}

\subsection{Magnetic Field Structure}\label{BGP} 

We derive the magnetic field structure of six filaments projected on the sky plane using the VGT, and the results are shown in Fig.\,\ref{para}\,and\,\ref{pendi}. 
The main skeleton of filaments which are identified by Python Package Radfil \citep{2018ApJ...864..152Z} and the results are also plotted in Fig.\,\ref{para} and\,\ref{pendi}.  
The sizes (lengths) of the filaments have been calculated by distance parameter from the BeSSeL Survey \citep{2019ApJ...885..131R}.

In this study, we utilize the VGT technique to measure the magnetic field, with $^{13}$CO map. 
Unlike dense gas tracers, such as, C$^{18}$O and NH$_3$, $^{13}$CO specifically traces the diffuse layer gas within giant filaments \citep{2019ApJ...884..137H,2022ApJ...934...45Z}. 
By focusing on this diffuse layer gas in VGT analysis, we obtain a better understanding of the magnetic field at larger scales \citep{2022ApJ...941...92H,2023MNRAS.519.1068L}, as it could be tracers of Galactic dynamics and less affected by the local physical process, such as star formation activities \citep{2019ApJ...884..137H}. 
Despite the limitations posed by the diffuse gas tracer and low spatial resolution in our study, the impact on the derived large-scale magnetic field through VGT is minimal, considering that the high column density regions occupy a small fraction of the total space.

Due to the fact map of the magnetic field has lower resolutions compared to the original map, we compare magnetic field orientation with the Galactic plane direction along the filament skeleton to study the relation between the magnetic field and the Galactic Plane at the filament body.
The Galactic Plane, physical mid-plane of Milky, is located 25 pc below the Solar \citep{2014ApJ...797...53G} and 7 pc above the Galactic center (Sgr\,A)\citep{2004ApJ...616..872R} (see Fig.\,\ref{figGP}. 
The offset angle between magnetic field orientation and the Galactic plane has been calculated by removing the offset angle between the physical mid-plane of the Galaxy and the IAU-defined mid-plane (b = 0\,$^\circ$; see the Fig.\,\ref{para} in \citealt{2014ApJ...797...53G}):
\begin{equation}
    \begin{aligned}
        \rm \theta_{B-GP} = \rm \theta_B - \rm \theta_{GP} ,
    \end{aligned}
\end{equation}
where $\theta_{\rm B}$ is magnetic field orientation.
$\theta_{\rm GP}$ is orientation angle of Galactic Plane ($\theta_{\rm GP}$ = b\,-\,0.12$^\circ$ (see Fig.\,\ref{figGP}; detail of $\theta_{\rm GP}$ shown in Tab.\,\ref{tab1}).
When the offset angles $\theta_{\rm B-GP}$ are less than $\pm$ 30$^\circ$, the magnetic field is near parallel to the Galactic Plane.
If the absolute value of $\theta_{\rm B-GP}$ is larger than 60$^\circ$, the magnetic field is close to being perpendicular to the Galactic Plane.
As shown in the Ap.\ref{apB}, the alignment Measure is the parameter to study the difference between two vectors.
When the alignment measures are above 0.5, the two vectors are considered parallel alignment \citep{2019NatAs...3..776H,2019ApJ...884..137H,2021ApJ...912....2H,2021MNRAS.502.1768H,2022ApJ...934...45Z}.
The alignment measure of 0.5 is equivalent to the offset angle between two vectors as 30$^\circ$ (see eq.\,\ref{AMeq}).
The Alignment Measure between VGT and other direction $\textgreater$ 0.5 is thought of as parallel alignment, where the offset angle between VGT and other orientation is less than 30$^\circ$ (see Ap.\ref{apB}).
Based on these reasons, we propose a threshold of 30$^\circ$ measured between VGT B-field and Galactic disk mid-Plane as the threshold below which the filament is considered as parallel. 
The distributions, along the skeleton of filament, of the offset angle $\theta_{\rm B-GP}$ between magnetic field orientation and Galactic Plane, are shown as in Fig.\,\ref{para}. and \ref{pendi}, where show the magnetic field distribution at the large scale of filament's main part.
The values of most offset angles $\theta_{\rm B-GP}$ are less than $\pm$ 30$^\circ$ and the magnetic field at the filament body is almost parallel to the Galactic Plane.
We use a new parameter, the fraction of effective area on the filament body p$_{\rm para}$ to show the percentage of parallel alignment between their magnetic field and Galactic Plane in filament body:
\begin{equation}
    \rm p_{para} = A_{para} / A_{filament}
\end{equation}
where the A$_{\rm para}$ is the effective area on the filament body where magnetic field orientations are parallel to the Galactic Plane, and A$_{\rm filament}$ is the total area of the filament body.
Because the B-field beam size is comparable to the filament width, the parallel fraction p$_{\rm para}$ can be estimated as the ratio between the normalized length L$_{\rm normalize, parallel}$ and the total length L$_{\rm normalize, total}$ (see Fig.\,\ref{para}, and Fig.\ref{pendi}).
The parallel probability p$_{\rm para}$ in most filaments is above 70$\%$ and that in the other two filaments, G339 and G349, is also above 50$\%$. Our results indicate that the magnetic field orientation at these filaments is regular and its direction stays mostly parallel to the Galactic plane at most filaments in our sample.
In some regions of filament G24, G339, and G349, the magnetic field appears to be discontinuous. These discontinuities are called "Magnetization Gaps".

\section{Discussion}\label{discu}

\subsection{Physical Properties and Galactic Distributions}

The six giant filaments in our sample are distributed at the inner Galaxy ($\sim$ 4-8\,kpc, see Fig.\,\ref{GF}). 
The Galactic scale height at the inner Galaxy is around 40\,pc \citep{2006PASJ...58..847N}. 
These filaments in our sample have a similar mass (10$^4$ $\sim$ 10$^5$ M$_{\sun}$), which is comparable to 
some of the massive molecular clouds\citep{2020MNRAS.493..351C}, have lengths that are larger than the 
scale-height of the Galactic disk. 
The resolution of the B-field probed by the VGT towards our sample sources ranges from 1 to 10 pc. 
The average H$_2$ column density is around 10$^{22}$ cm$^{-2}$. 





\begin{figure}
    \centering
    \includegraphics[width =8cm]{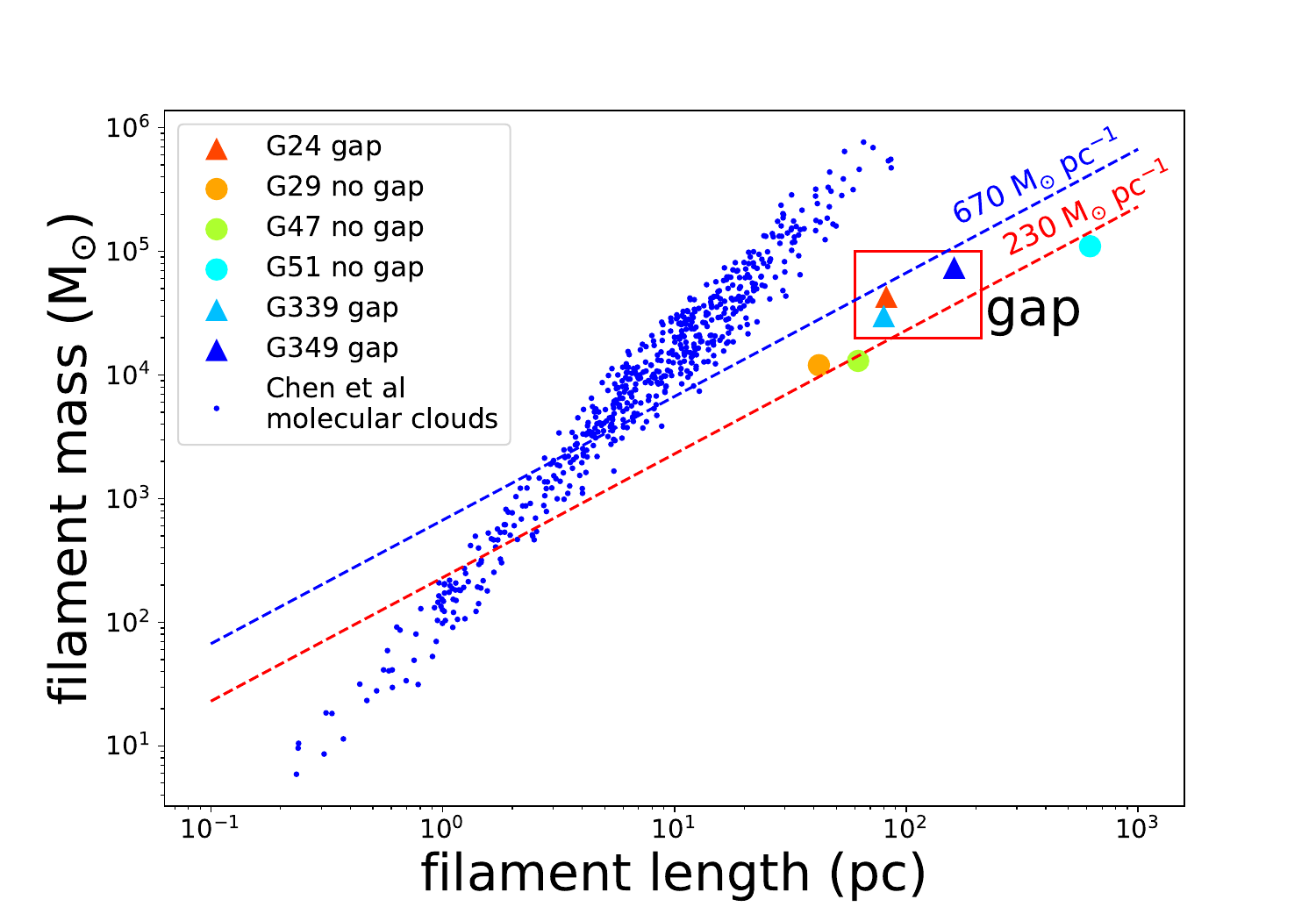}
    \caption{\textbf{Filaments and molecular clouds in the mass-size plane}. 
    The triangles show the filament with magnetization gaps in the red frames. 
    The circles show the filaments (no gap). The red and blue dot lines are 230 and 670 M$\sun$ pc$^{-1}$ which are representative line masses of non-star-forming and star-forming filaments \citep{2016A&A...591A...5L}. 
    The blue dots represent the molecular cloud from \citealt{2020MNRAS.493..351C}.}
    \label{linemass}
\end{figure}

\begin{figure}
    \centering
    \includegraphics[width = 8.6cm]{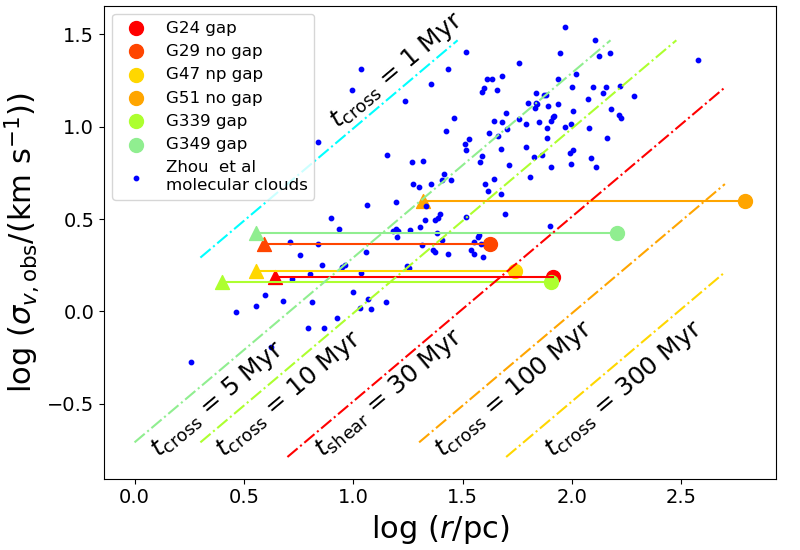}
    \caption{{\bf Distribution of our filaments in the velocity-size plane.}
    The circles show the crossing time of giant filaments, where the detail of cross time is shown in Tab.\,\ref{tab1}. 
    The triangle displays the crossing time measured with the resolution of our VGT calculations as the characteristic length (see Eq.\,\ref{eq13}, where the detail of B-field cross time is shown in the Tab.\,\ref{tab1}. 
    The blue points represent a sample of YSO complex \citet{2022MNRAS.513..638Z} where we use the motion of the young stars to study the turbulent motion of the molecular gas.}
    \label{figtcross}
\end{figure}

\begin{figure*}
    \centering
    \includegraphics[width = 14cm]{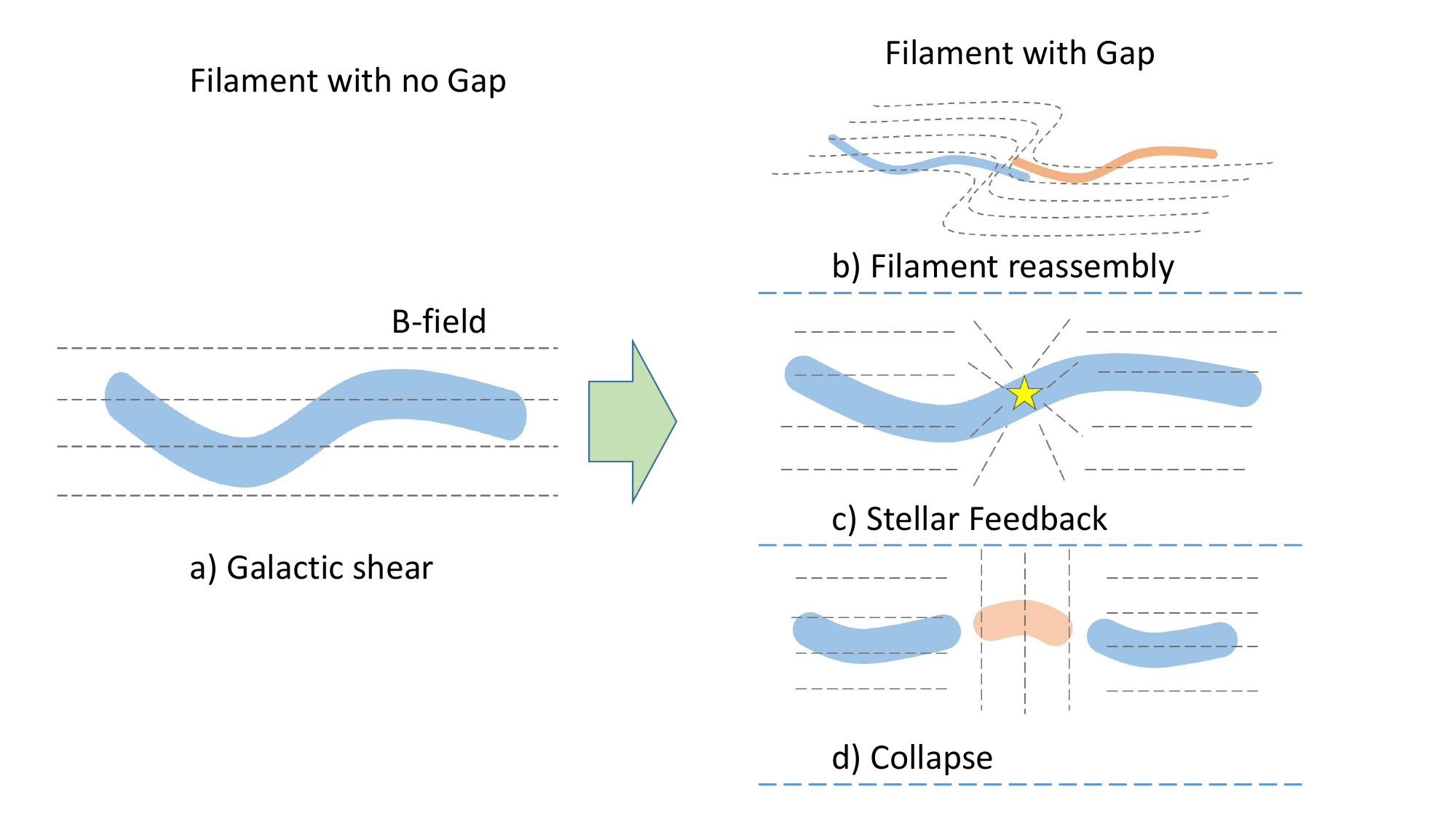}
    \caption{\textbf{Possible mechanisms leading to the production of magnetization gaps, which include: filament reassembly\,(b), stellar feedback\,(c), and gravitational collapse\,(d)}. The dot black lines show the B-field. The blue and orange lines show the shape of filaments. The yellow star represents star formation. (See Sec.\,\ref{sec5.4} for details.)}
    \label{filament-B}
\end{figure*}

\subsection{Shear-induced Magnetic Field-filament Alignment}\label{sec5.2}

As shown in Fig.\,\ref{para} and \ref{pendi}, for all filaments (G24, G29, G47, G51, G339, and G349, see Appendix\,\ref{Ap.C}), the magnetic field stays aligned with the filament bodies. 
Since the filament is already aligned with the midplane of the Galactic disk \citep{2016ApJS..226....9W,2016A&A...591A...5L,2022MNRAS.513..638Z}, the B-field, which tends to stay parallel to the filament body, also stays parallel to the disk mid-plane.

VGT method is capable of revealing the magnetic field morphology using spectrally-resolved maps. 
The magnetic field measured by VGT is a two-dimensional representation of the true field. 
Although it removes the influence of the background and foreground, this projection effect is still a severe limitation.
Magnetic field orientation measured by VGT is inevitably affected by the projection effect. 
Nevertheless, the alignment between the magnetic and the filament body can be retained even if we view the filament in projection. 

One possible explanation for the alignment between the magnetic field orientation is Galactic shear. In this scenario, shear-strength gas is into long filaments, and this very process creates a magnetic field that is parallel to the filament body and the disk midplane. This scenario is shown in Fig.\,\ref{filament-B}.
\citet{2022MNRAS.513..638Z} estimate a lifetime of $\gtrsim $60 Myr for cloud-sized YSO associations. 
In nearby galaxies, the Giant molecular clouds have an estimated lifetime of 10 $\sim$ 30 Myr \citep{2020MNRAS.493.2872C}. 
If our filaments have a similar lifetime, their orientation can be determined by shear \citep{2013A&A...559A..34L,2016A&A...591A...5L,2022MNRAS.513..638Z}.





\subsection{Regular magnetic field at $\approx$ 10 pc scale: Stability against turbulent perturbations} \label{sec5.3}

There are a few timescales controlling the evolution of the clouds. The fact that the magnetic field stays regular and parallel to the mid-plane of the Galactic disk mid-plane suggests that the shear plays a role in inducing this alignment, yet by evaluating these timescales, one can constrain their roles in 
the these processes in cloud evolution. These timescales include the shear time
\begin{equation}
    \it{t}_{\rm shear} = \it{\kappa}^{-1} = (\rm 1/\it{r}|{\rm d}{\rm \Omega} /d\it{r}| = (\rm 2\it{A})^{-1} = \rm 30\, Myr \;,
\end{equation}
where $\kappa$ is the shear rate, A is the Oort constant ($\approx$ 16 km\,s$^{-1}$\,kpc$^{-1}$;\citealt{2021MNRAS.504..199W}) in the Solar neighborhood.
The cross-time of giant filaments is shown in Fig.\,\ref{figtcross} determined using:
\begin{equation}
    t_{\rm cross} = L / \sigma_v
\end{equation}
where L is the length of the filament's long axis, $\sigma_v$ is the velocity dispersion of $^{13}$CO spectral lines. 
These cross time of filaments are mostly above the lifetime of Giant molecular clouds (10$\sim$30 Myr; \citealt{2020MNRAS.493.2872C}).
These filament structures could be shaped by galactic rather than local turbulence.

The resolution of the magnetic field in our sample has regular structures at around 10 pc scale, which is close to the width of the filament.
The time, turbulence crossing B-field resolution size is calculated as:
\begin{equation}\label{eq13}
    t_{\rm B,cross} = L_B / \sigma_v
\end{equation}
where L$_B$ is the resolution of magnetic field in our samples, $\sigma_v$ is the velocity dispersion of $^{13}$CO spectral lines. 
As shown in Fig. \ref{figtcross}, this crossing time, referred to as the B-field cross, is around 1 to 5 Myr.
The B-field crossing time is below the shear time ($\sim$ 30 Myr). Within the timescale of shear, the magnetic field exhibits regular structures and stability against turbulent perturbations, and this is only possible if the magnetic field stays unchanged at the resolution scale of around 2.5 - 10\,pc. Thus, turbulent motion at the small scale is ineffective in disrupting the magnetic field at the large scale.

\subsection{Magnetization Gaps: Stellar Feedback or Filament Assembly}\label{sec5.4}

Magnetization gaps are regions where the magnetic field appears to be discontinuous. These magnetization gas are often accompanied by gaps in the surface density distribution of the gas. Examples including  G24, G339, and G349, (see Fig.\,\ref{pendi}). 
Since the estimated magnetic field orientation stays parallel to the disk mid-plane, the gaps are likely caused by some additional forces. Here are some possibilities:

\subsubsection{Filament Reassembly}


The magnetization gaps may represent remnants of the filament reassembly process. This phenomenon occurs when segments of distinct filaments amalgamate to form a new filament.
A particularly intriguing illustration can be found in the case of filament G339, where multiple velocity components, each separated by a few kilometers per second, are discernible within its magnetic gap (shown in Fig. \ref{pendi}).
These various velocity components can be attributed to the filament reassembly process, a phenomenon demonstrated in simulations presented in \citealt{2013MNRAS.432..653D}.
Before the filament reassembly, the magnetic field was continuous and aligned with the mid-disk plane, as discussed in Sect. \ref{sec5.2}.
However, the filament reassembly resulted in the merging of magnetic fields from different filaments, rendering the magnetic field discontinuous and giving rise to the magnetization map (See Fig. \ref{filament-B}).

One supporting evidence is that these ''gap filaments'' -- filaments with magnetization gaps are distributed in the Norma-Outer and Scutum-Centaurus-OSC arms, which are close to the Galactic Center (see Fig.\,\ref{GF}),  with the Galactocentric distances of the three gap filaments, stay within 6\,kpc. 
At a smaller Galactocentric distance, the disk is crowded with clouds, and the shear time is shorter. 
These factors should increase the chances for structures to reorganize \citep{2017MNRAS.471.2002L,2022arXiv220805303L}, leading to a higher fraction of filaments with magnetization gaps. This scenario is illustrated in Fig.\,\ref{filament-B}.

\subsubsection{Collapse and Stellar Feedback}
We propose that collapse and stellar feedback can be the cause of the magnetization gaps. 

Collapse tends to occur in localized regions of sizes of $\gtrsim \;   1 \,\rm pc$  \citep{1981MNRAS.194..809L,2017MNRAS.465..667L}. 
The collapse, which is gravity-dominated,  can lead to localized contractions, which in turn induce the formation of discontinuities produced in this process that can be recognized as gaps (see Fig.\,\ref{filament-B}).
In filament G24 (see Fig.\,\ref{pendi}), a dense clump identified by ALTASGAL \citep{2014MNRAS.443.1555U} is located at the position of the magnetized gaps. 
This cluster comprises Young Stellar Objects (YSOs), indicating that the gravitational condensation of gas results in the creation of new stars. The processes of collapse and star formation may distort the nearby large-scale magnetic field, causing it to transition from a state of continuity to discontinuity, thereby giving rise to magnetization gaps.
The magnetization gap on the large scale could potentially serve as an indicator for identifying the locations where the collapse or initiation of star formation activities transpire at the sub-beam size.

As shown in Fig.\,\ref{filament-B}, another possibility is that these gaps are caused by stellar feedback. 
Stellar feedback occurs when massive stars disrupt nearby clouds through stellar wind, radiation pressure, and supernova. 
The very process of cloud disruption can push the gas around, leading to distortions of the magnetic fields.
These distortions can appear as, leading to a magnetization gaps magnetization gap. 
As Fig.\,\ref{pendi} shows, a dense clump in filament G349, identified by ALTASGAL \citep{2014MNRAS.443.1555U}, is located at the position of the magnetized gap. \citealt{2014MNRAS.443.1555U} consider this clump as an HII region, providing evidence of stellar feedback impacting the magnetized gap.

We note that as shown in Fig.\,\ref{linemass}, the line mass of filaments with gaps is higher than that of filaments without gaps, which is consistent with either scenario.
The line mass of the filaments with magnetization gap is above the 230 $M_\odot$ pc$^{-1}$, which is on the higher side  \citep{2016A&A...591A...5L}. 
It is possible that higher line mass leads to a faster collapse which disrupts the fields. The subsequent star formation actives can also disrupt the filaments, leading to gaps. 
The different possible mechanisms of magnetization gaps are illustrated in Fig.\,\ref{filament-B}.

\section{Conclusion}\label{sum}

We study the structure of magnetic fields in six giant filaments probed by the VGT technique.
The main results include:

In most of our sample objects, the magnetic field orientation appears to be aligned with the filament which stays parallel to the mid-plane of the Galactic disk. 
The alignment between the magnetic field, filament, and mid-plane of the Galactic disk could be the result of the Galactic shear. 
At the magnetic field around 10 pc scale, the magnetic field maintains the regular structure. 

Magnetization gaps are detected towards filaments G24, G339, and G349. 
These are discontinuities of inferred regular magnetic field structures and are easily visible from the maps, Meaning that turbulent motion can not effectively disrupt the magnetic field structure. On the other hand, the filament reassembly process, gravitational collapse, and stellar feedback could affect the magnetic field of giant filaments, leading to these magnetized gaps. 

Further studies towards larger samples would provide a clearer picture of B-field-filament alignment and reveal the origins of the magnetization gaps.


\section*{ACKNOWLEDGEMENTS}

We thank Yue Hu and Prof. Alex Lazarian for VGT code and helpful comments. This work acknowledges the support of the National Key R$\And$D Program of China under grant No. 2022YFA1603103, the National Natural Science Foundation of China (NSFC) under grant Nos. W820301904265, 12033005, 11973076, 12173075, and 12103082,12303027, the Tianshan Talent Program of Xinjiang Uygur Autonomous Region under grant No. 2022TSYCLJ0005, the Natural Science Foundation of Xinjiang Uygur Autonomous Region under grant No. 2022D01E06, the Chinese Academy of Sciences (CAS) “Light of West China” Program under grant Nos. xbzg-zdsys-202212, 2020-XBQNXZ-017, and 2021-XBQNXZ-028, the Xinjiang Key Laboratory of Radio Astrophysics (No. 2023D04033), the Youth Innovation Promotion Association CAS, the Regional Collaborative Innovation Project of Xinjiang Uyghur Autonomous Region (2022E01050), the Project of Xinjiang Uygur Autonomous Region of China for Flexibly Fetching in Upscale Talents, and the Hebei NSF No. A2022109001.

\bibliography{reference}

\appendix
\section{Magnetic field morphology}\label{Ap.C}

\begin{figure*}
    \centering
    \includegraphics[height = 5.8cm]{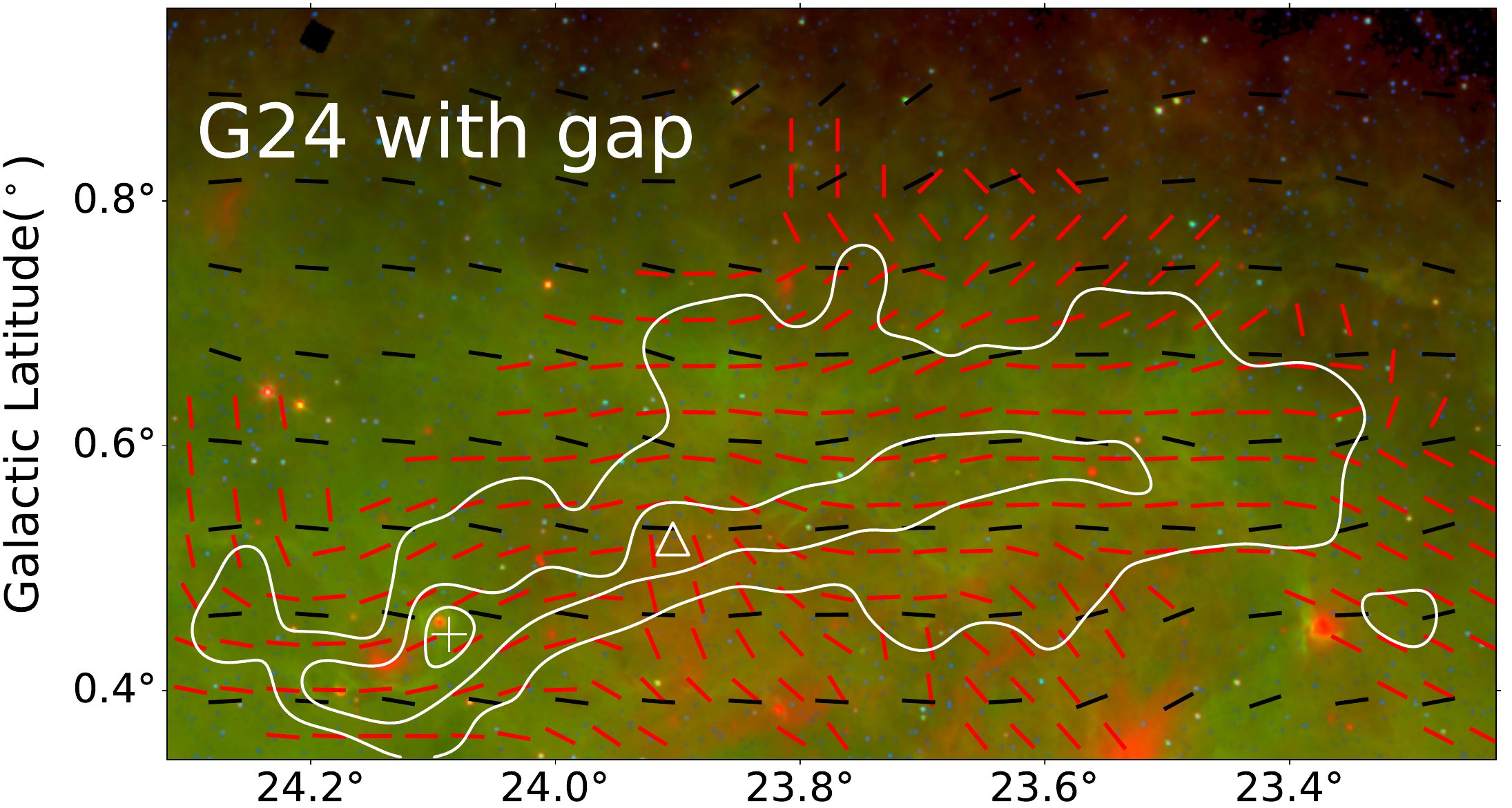}
    \includegraphics[width = 18cm]{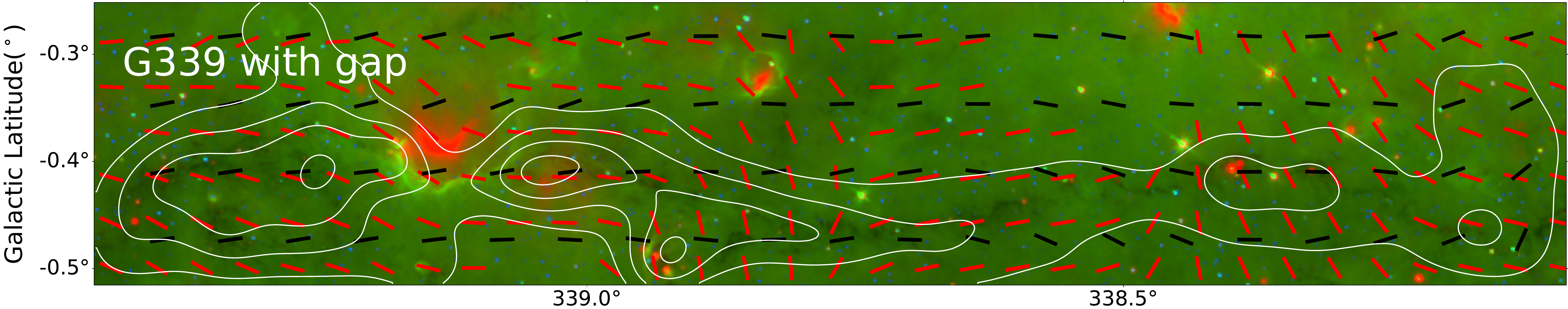}
    \includegraphics[width = 18cm]{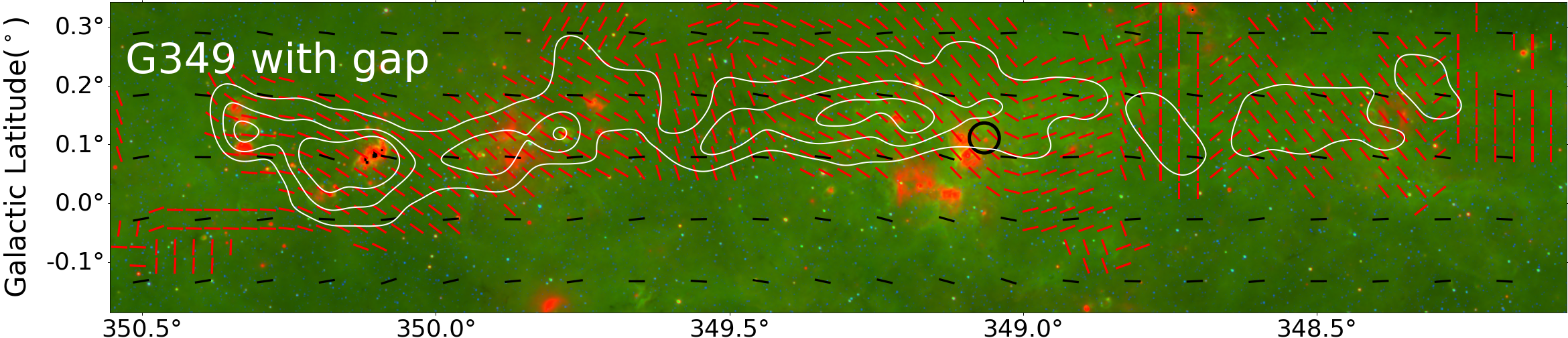}
    \includegraphics[height = 4.4cm]{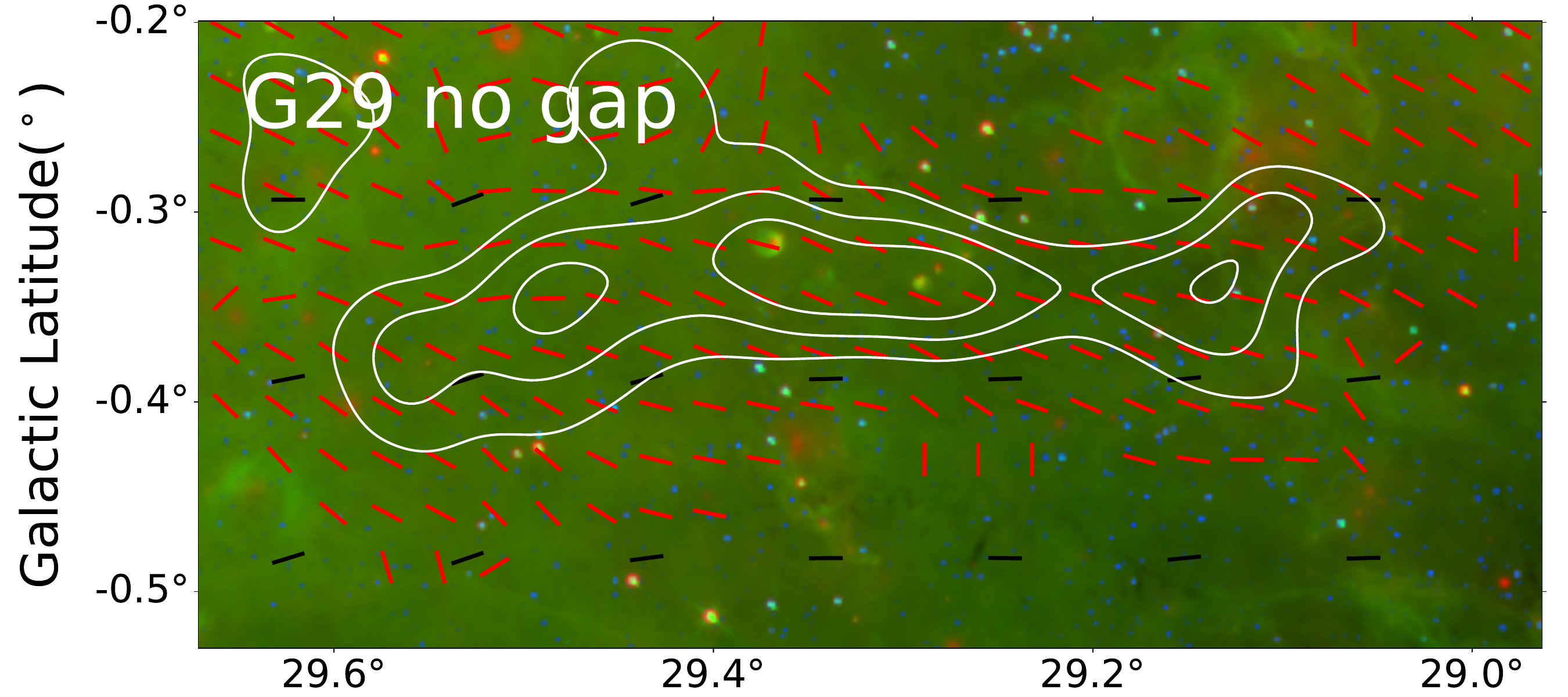}
    \includegraphics[height = 4.4cm]{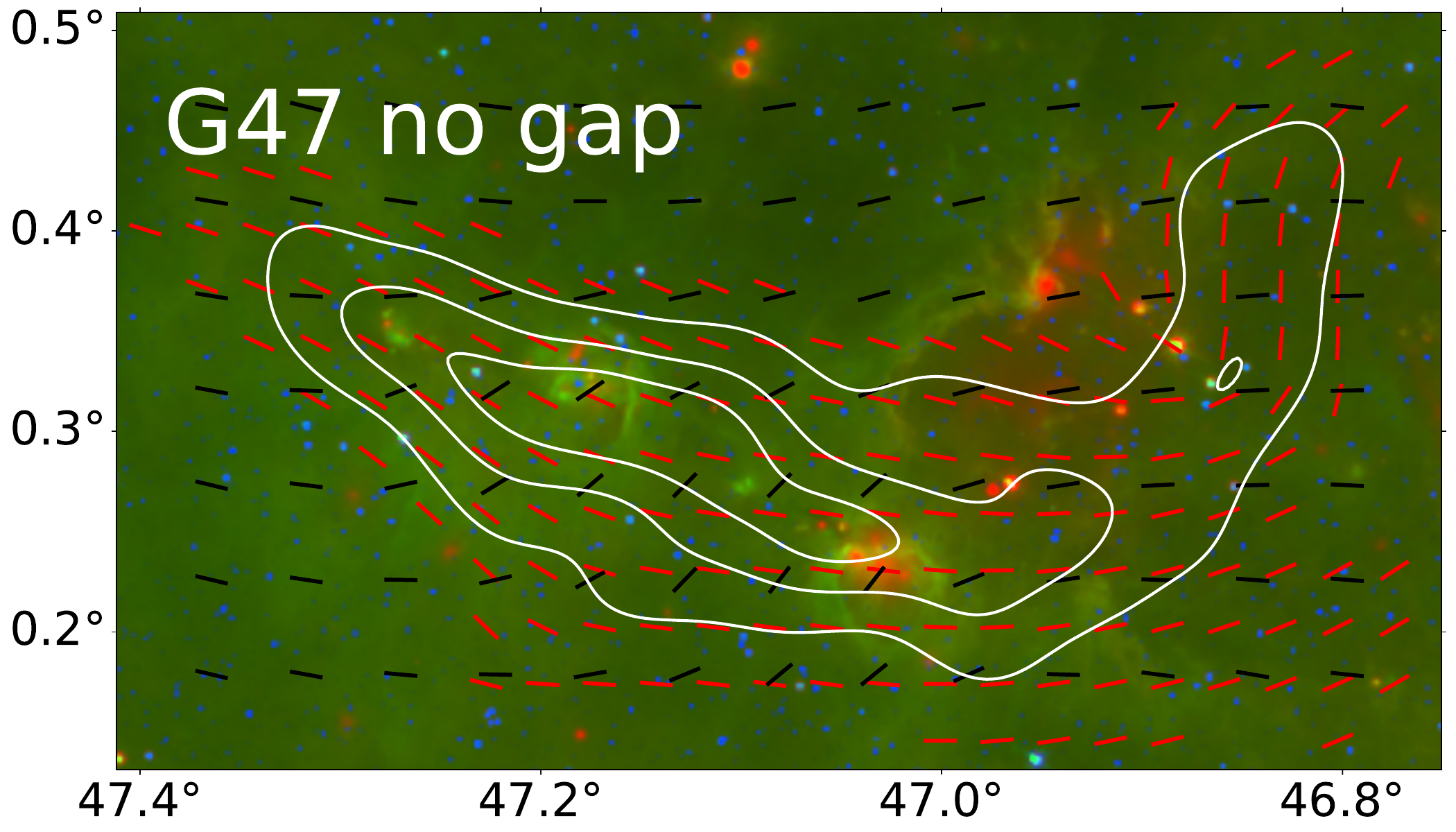}
    \includegraphics[width = 18cm]{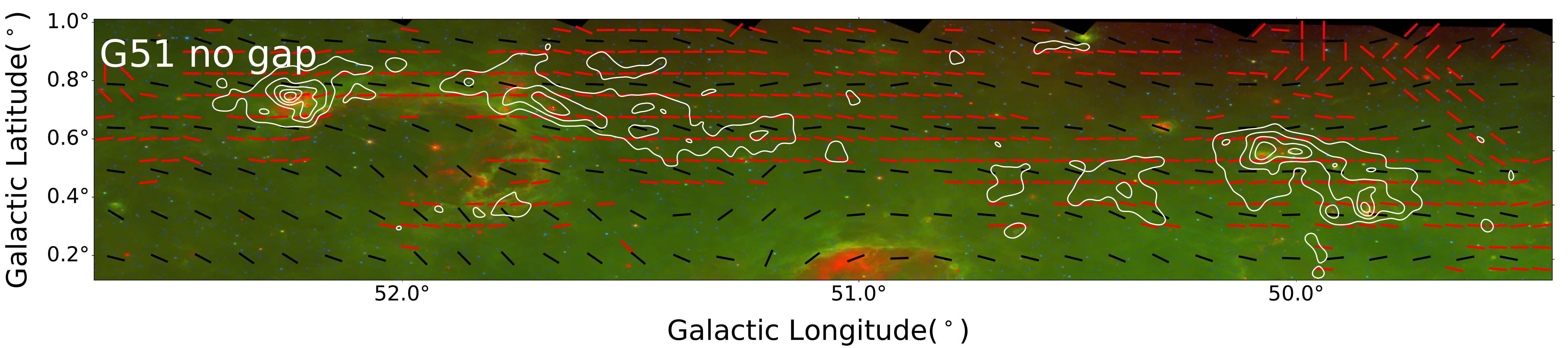}
    
    \caption{\textbf{Mid-IR RGB image of six filaments}. In $\it{Spitzer}$ (GLIMPSE \citep{2003PASP..115..953B} and  MIPSGAL \citep{2009PASP..121...76C}) three-color image, red is 24$\mu$m; green is 8$\mu$m; blue is the 3.6$\mu$m. The red vectors show the magnetic field orientations derived by the VGT method for $^{13}$CO and the Black line shows the Planck orientation. The white contour shows the shape of $^{13}$CO intensity map of filaments. }
    \label{C1}
\end{figure*}

As shown in Fig.\,\ref{para} and \ref{pendi}, the magnetic field structure of our sample has been shown as line-integral-convolution (LIC) map\citep{Cabral_lic}. 
Star formation activity occurs in the filament. 
24\,$\mu$m emission could trace the star formation activity, where dust could be heated by the newly-born stars.
These regions are shown as the red area in the $\it{Spitzer}$ image (see Fig.\ref{C1}).
In Fig.\,\ref{C1}, we found that magnetic field orientations measured with VGT (red vectors) have a similar trend to that derived by Planck 353GHz dust polarization (black vectors) at most parts of the sample.
The B-field derived by Planck uses the dust polarization at 850\,$\mu$m continuum. 
The foreground and background emissions largely affect the B-field derived by Planck. 
By comparison, the magnetic field measured by the VGT method for $^{13}$CO could be a cleanly local B-field of our sample in the low Galactic Latitude region.
The magnetic field orientations of six Galactic-scale filaments, measured by the VGT method for $^{13}$CO emission, are shown as follows:

\noindent\textbf{\textcolor{blue}{G24 with gap}}. 
G24 is an 'L' shape like filament
Magnetic field resolution is 2.83$^\circ$ (170$''$, subblock as 20$\times$20 pixels). 
Magnetic field orientation at the western filament is parallel to the filament direction. 
At the eastern filament, there are two density clumps (cross: G24.087+0.446, triangle: G23.904+0.524). G24.087+0.446 is the densest region at the G24 filament. Magnetic field orientations tend to converge towards this Clump. G23.904+0.524 is another density clump at this filament.
There are much emission from 24\,$\mu$m (see Fig.\,\ref{C1}). 
It means that there exists star formation activity.
Magnetic field directions are distorted and not parallel to the Galactic Plane. 
The star formation activity could distort the magnetic field structure and cause magnetization gaps.

\noindent\textbf{\textcolor{blue}{G29 no gap}}. 
There is a sine wave-shaped filament at G29.18-0.34.
The magnetic field orientation of G29 is roughly parallel to the filament direction. 
At the eastern and western of the filament, the magnetic field is parallel to the filament direction. 
At the filament westernmost, the filament shape is bifurcated. 
Magnetic field morphology exhibits distortion with the filament shape. 
The middle part of this filament is the densest area of this filament. 
The trend of the magnetic field direction is close to a slight deflection to the north and south. 

\noindent\textbf{\textcolor{blue}{G47 no gap}}. 
Filament G47 resembles a "C" shape. 
As the whole area of filament G47, magnetic field orientations are relatively consistent. 
It is close to parallel to the Galactic Plane.
In the eastern area, the B-field directions are parallel to the direction of the filament shape. 
In the western area, magnetic field orientations are parallel to filament shape but perpendicular to the Galactic Plane.
The magnetic field structure could be affected by the filament body.
In general, the magnetic field orientations are continuous and parallel to the Galactic plane and filament directions.

\noindent\textbf{\textcolor{blue}{G51 no gap}}. 
G51 is a giant filament. Its length has been over 500 pc \citep{2013A&A...559A..34L}. 
In Fig.\,2 bottom panels shown, two HII region, G52L nebula, and G50 bubble, is located south of the filament (see white lines). 
Magnetic field orientations around HII regions are parallel to the HII region shell direction. 
On the whole, magnetic field orientation is east-west direction and parallel to the filament direction.
This large-scale magnetic field structure could be affected by Galactic shear and the HII region around the filament.


\noindent\textbf{\textcolor{blue}{G339 with gap}}. 
Filament G339 (Nessie, \citealt{2014ApJ...797...53G,2015ApJ...815...23Z}) could be a part of the Milky Way spiral arms skeleton.
The magnetic field orientations are parallel to the Galactic Disk in most regions. 
There is a curve of the magnetic field at two clumps in the Nessie filament (see Fig.\,\ref{pendi} middle panel star, and cross). The magnetic field occurs discontinuously. 
As shown in Fig.\,\ref{pendi}, there could exist two velocity components in the magnetization gap region.
The P-V diagram shows that the two velocity components have a separation of around 3 to 4 km s$^{-1}$.
There is no the emission from continuum at 24\,$\mu$m and star-formation activity.
The filament reassembly could occur in this region and cause magnetization gaps.
At filament western, the magnetic field has suddenly changed where has the 24\,$\mu$m emission, and star formation.

\noindent\textbf{\textcolor{blue}{G349 with gap}}. 
Fig.\,\ref{pendi} bottom panel shows the filament G349 from 348$^\circ$ $\textgreater$ l $\textgreater$ 350.5$^\circ$. 
The background is intensity map of $^{13}$CO (2-1) emission integrated within v$_{lsr}$ as [-70, -56] km\,s$^{-1}$. 
The long axis direction of filament G349 is close to parallel to the Galactic Plane.  
The magnetic field orientation is northeast-southwest direction. 
There are two obvious magnetization gaps which are at G349.9+0.1 and  G348.7+0.1.
At the G349.9+0.1, there exists the star formation activity (see Fig.\,\ref{C1}).
It could cause magnetization gaps.
In another region, the filament could be fragmented and cause magnetization gaps.

\section{Comparison between VGT and Planck}\label{apB}

\begin{figure*}
    \centering
    \includegraphics[height = 5.8cm]{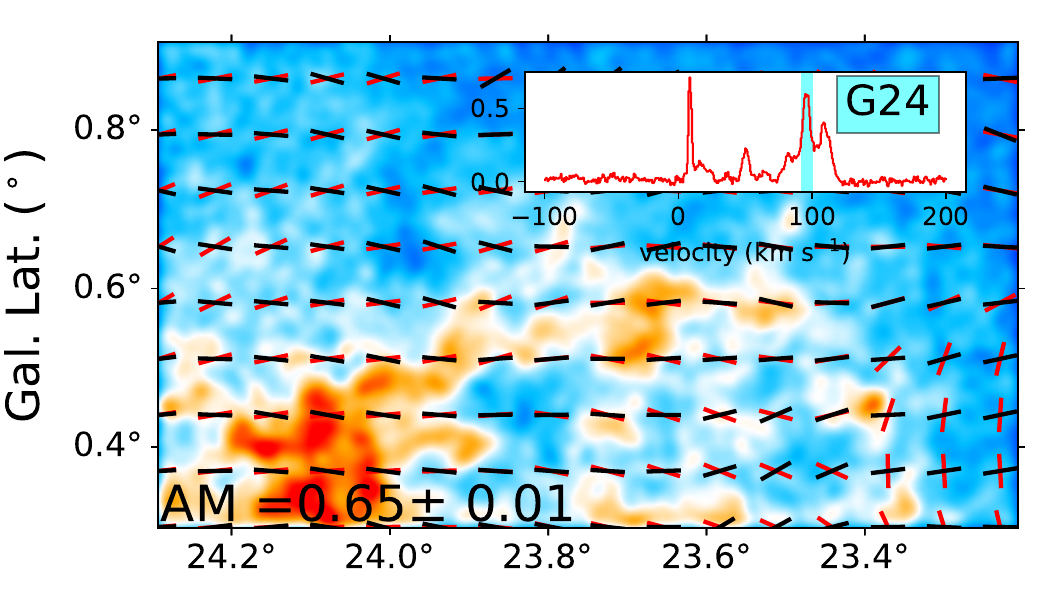}
    \includegraphics[width = 18cm]{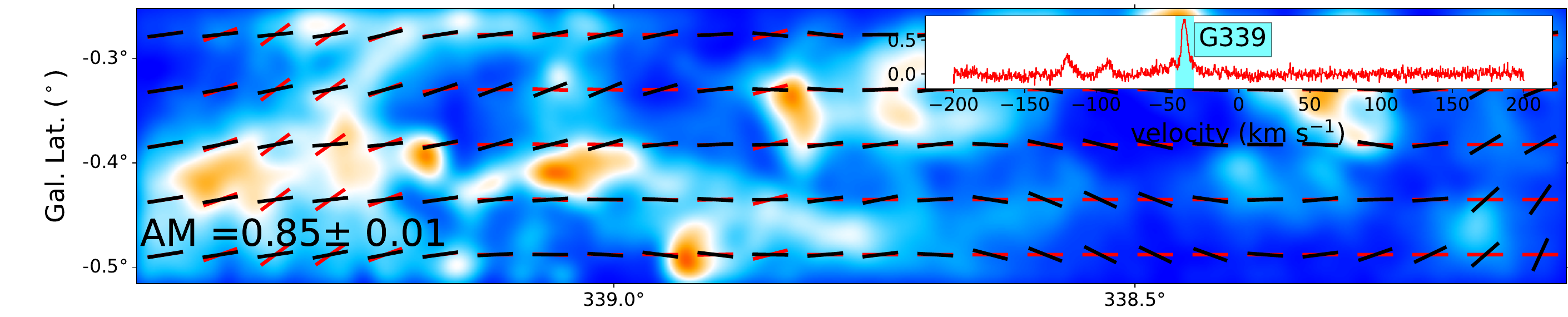}
    \includegraphics[width = 17.7cm]{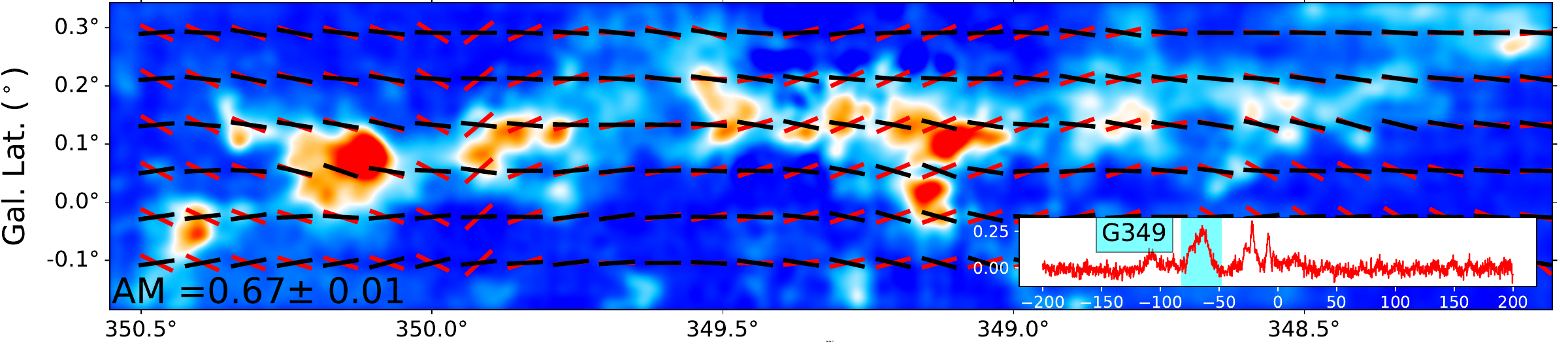}
    \includegraphics[height = 4.4cm]{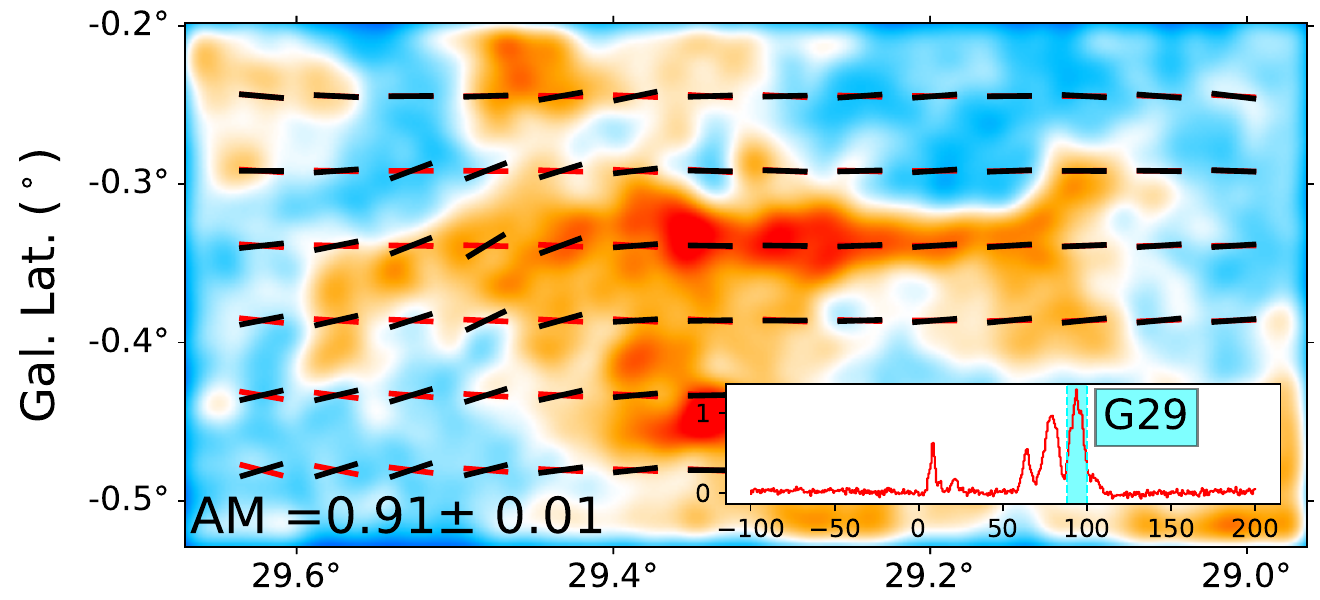}
    \includegraphics[height = 4.4cm]{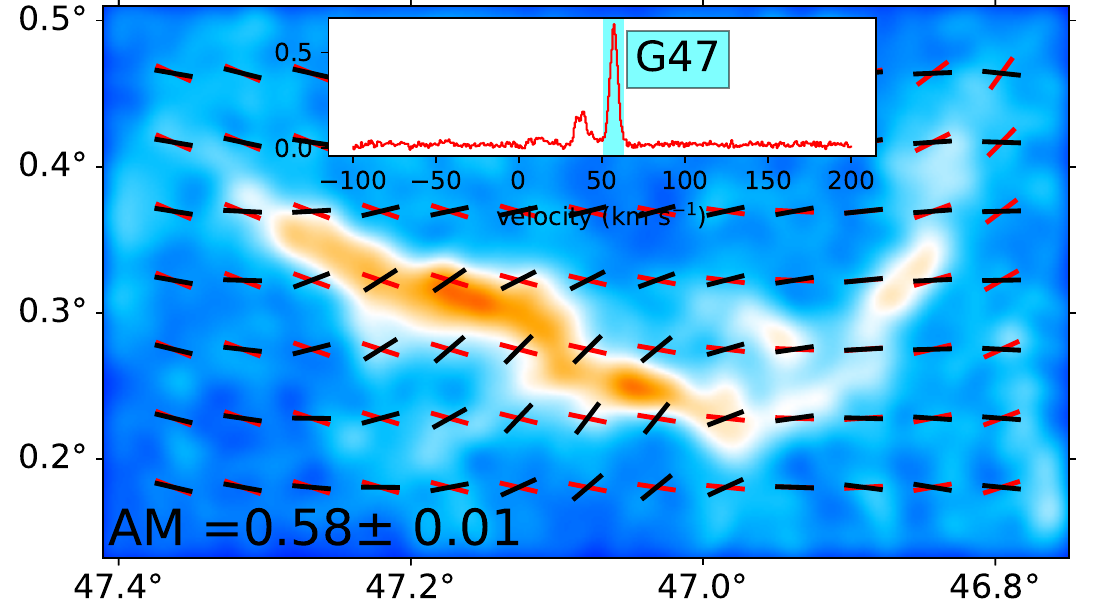}
    \includegraphics[width = 18cm]{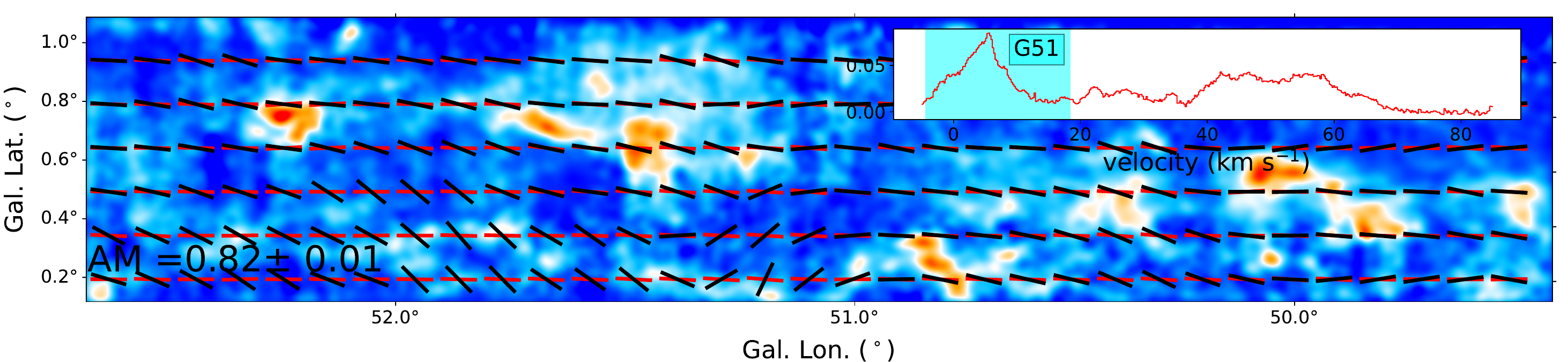}
    \caption{\textbf{Distribution of magnetic field measured with VGT for $^{13}$CO and dust polarization by Planck 353Ghz continuum.}
    The red vectors display the magnetic field derived by VGT for $^{13}$CO emission at the full velocity range.
    The black vectors show the magnetic field obtained by Planck 353GHz dust polarized continuum.
    The background is the intensity map of $^{13}$CO emission integrated with full velocity range.
    The mean AM value is shown in the left bottom corner of each panel. 
    The sub-panel displays the mean spectral line and the cyan lines show the velocity range of giant filaments.}
    \label{CVB}
\end{figure*}

To show the advance of VGT tracing magnetic field with multiple velocity components, we compare it with the Planck 353 GHz dust polarized continuums. 
Due to the dust-polarized continuums including the massive foreground in the Galactic disk, we integrate the spectral line on the VGT algorithm at the full velocity range to measure the magnetic field, which include the multiple velocity components in the LOS and their origins are similar to continuums.

For convenient comparison, we smooth the magnetic field measured by VGT to the same beam of Planck polarization. 
We smooth the VGT-field of G24, G29, G47, G339, and G348 to 5$'$ and smooth that of G51 to 10$'$. 
We use the Alignment Measure (AM, \citealt{2017ApJ...835...41G}) to study the difference between magnetic field derived by VGT and dust polarization:
\begin{equation}\label{AMeq}
    \begin{aligned}
        \rm AM = 2(\left \langle cos^2( |\phi_B - \psi_B^S|) \right \rangle - \frac{1}{2}) \,.
    \end{aligned}
\end{equation}
where the $\phi_B$ is magnetic field orientation derived by dust polarization and the $\psi_B^S$ is VGT-derived magnetic field orientation.
The range of AM values would be from -1 to 1, where  AM values are close to 1 mean that $\phi_B$ is parallel to $\psi_B^S$ and an AM value close to -1 indicates that $\phi_B$ is perpendicular to $\psi_B^S$. 
The uncertainty in the AM value, $\sigma_{AM}$, may be given by a standard deviation divided by the square root of the sample size.

As Fig.\,\ref{CVB} shows, the mean AM of filaments is over 0.5. 
It means that the magnetic field measured with VGT is close to being parallel to the orientation of dust polarization. 
Weak differences still exist between the two types of techniques measuring magnetic fields.
There exists a possibility.
The depth of spectral line integration is not deep enough. 
The origin of the spectral line is a little different from that of Planck continuums in order to the weak difference between the two types of the magnetic field \citep{2022ApJ...934...45Z}.
Another possibility is that the integrated depth of spectral survey also affects the AM between VGT and Planck. 
The low integrated depth causes the low mass region not be reflected at the spectral lines.
Meanwhile, this also results in the VGT for full velocity range and Planck dust polarization origin from different region and their magnetic field orientation exist difference.

As Fig.\,\ref{CVB} shows, the giant filaments have massive foreground. 
Using VGT to measure the magnetic field, we only extract the effective velocity component, which is the local velocity range of filaments as shown in Tab.\,\ref{tab1} and Fig.\,\ref{CVB}.
By extracting the effective velocity component in spectral lines, the magnetic field measured with VGT for spectral PPV cube can avoid the effect of foreground and background.
It is an obvious advantage of VGT compared with other measurements, which can clearly measure the magnetic field of giant molecular clouds with other velocity components in LOS.

\section{Calculating the Angle Between Objects and Galactic Plane}

\begin{figure}
    \centering
    \includegraphics[width = 17cm]{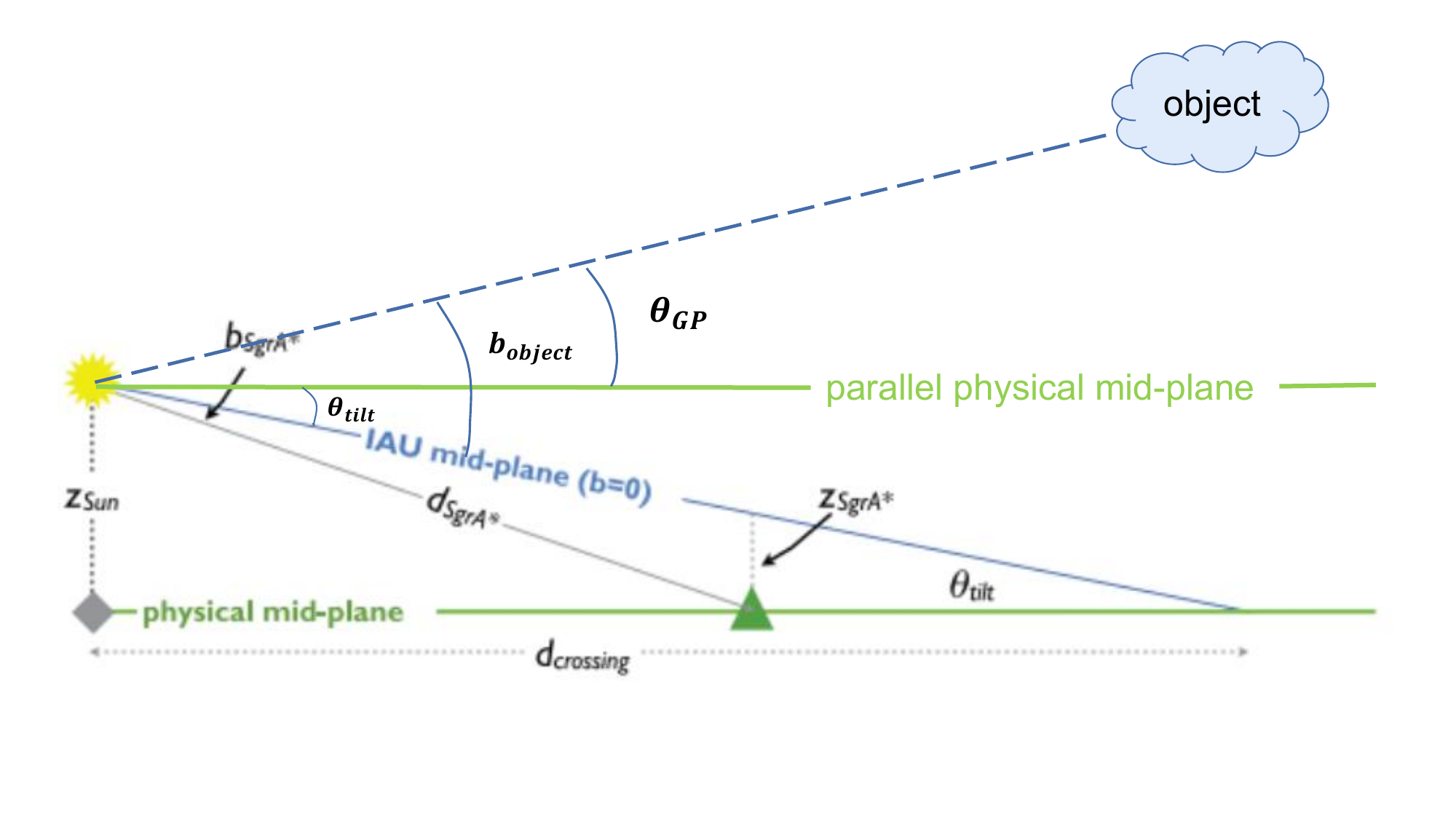}
    \caption{\textbf{Calculating the Angle Between Objects and Galactic Plane.}
    The caption is sourced from \citealt{2014ApJ...797...53G}.
    The physical mid-plane is situated 25 pc below the Solar and 7 pc above the Galactic center (Sgr\,A).
    According to \citealt{2014ApJ...797...53G}, they discovered that the angle between the physical mid-plane (Galactic Plane) and the IAU mid-plane (b = 0$^\circ$), denoted as $\theta_{tilt}$, is approximately 0.12$^\circ$.
    In the Galactic Coordinate System, the parameter 'b' represents the angle between an object and the IAU mid-plane (b = 0$^\circ$).
    The angle between the object and the Galactic Plane (physical mid-plane), denoted as $\theta_{GP}$, is calculated as b - $\theta_{tilt}$ (b - 0.12$^\circ$).}
    \label{figGP}
\end{figure}

\section{Position Velocity Diagram along Filament Long axis}

\begin{figure}
    \centering
    \includegraphics[width = 17cm]{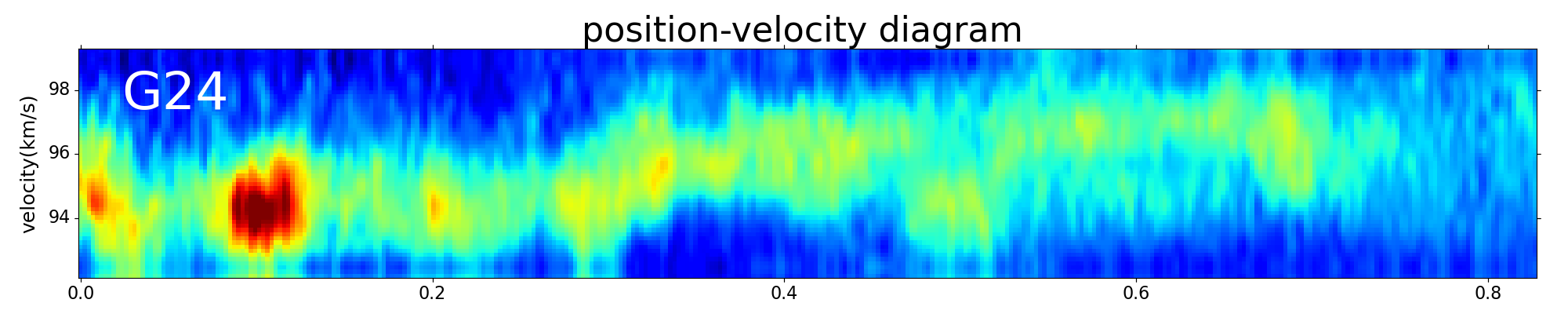}
    \includegraphics[width = 17cm]{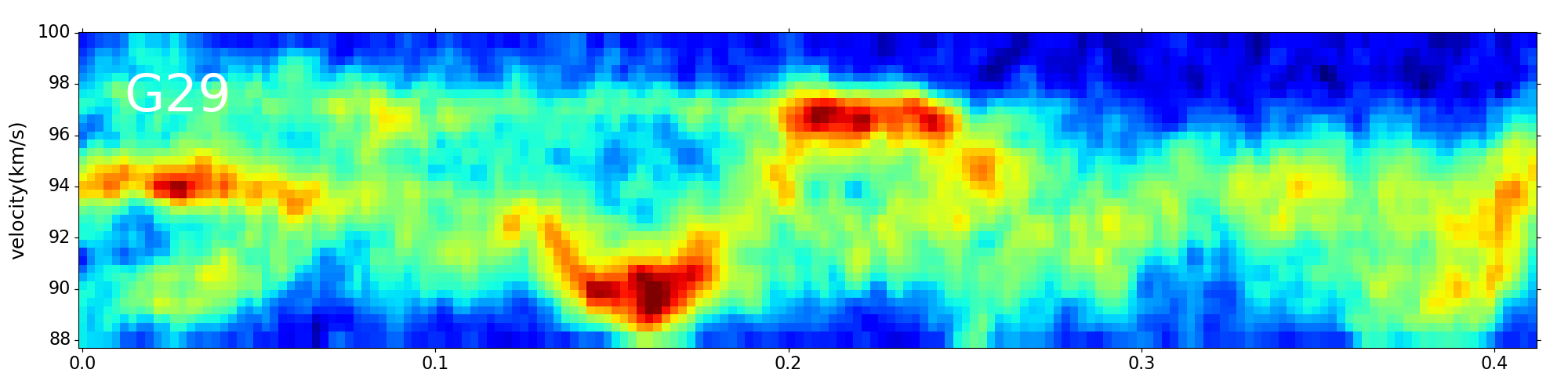}
    \includegraphics[width = 17cm]{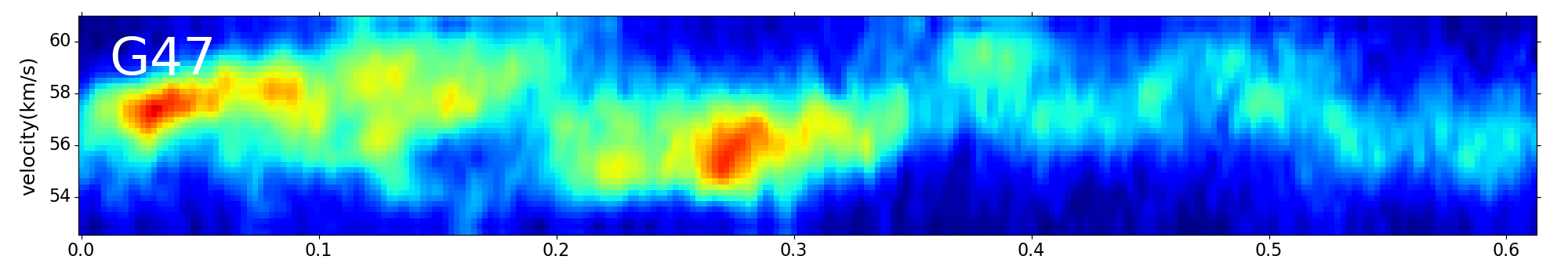}
    \includegraphics[width = 17cm]{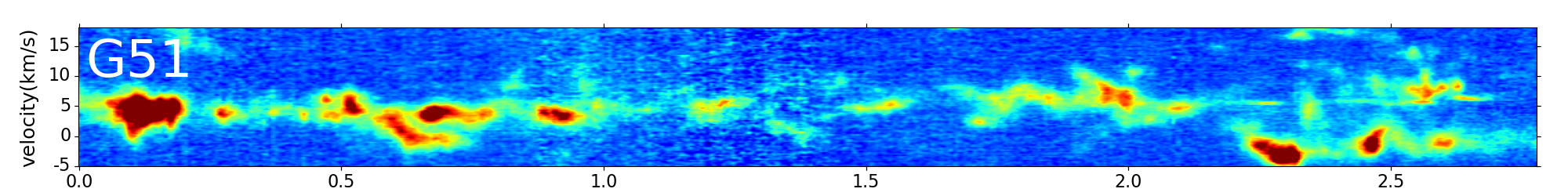}
    \includegraphics[width = 17cm]{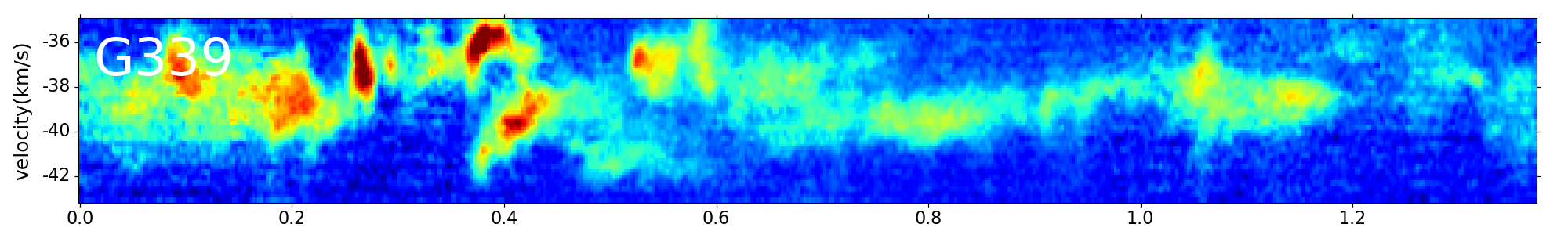}
    \includegraphics[width = 17cm]{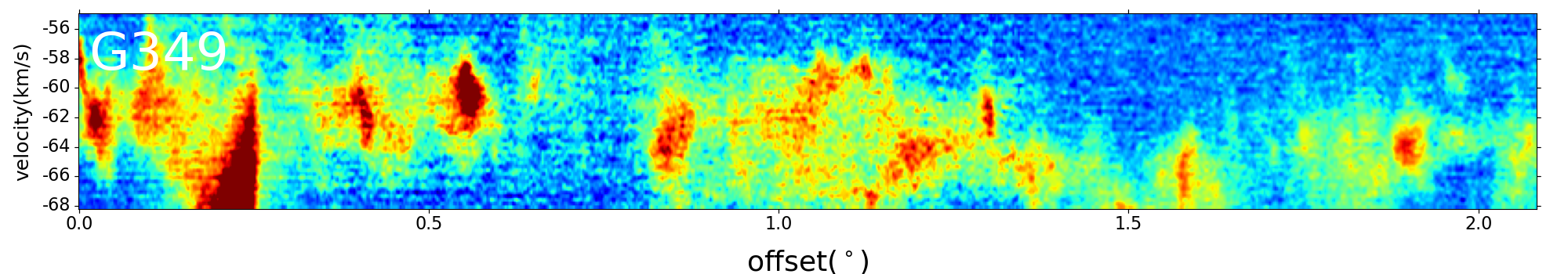}
    \caption{\textbf{The position velocity diagram along filament long axis.}
    These panels show the P-V diagram for six giant filaments along the filaments' long axis which is the same as Fig.\,\ref{para}, and Fig.\,\ref{pendi}.}
    \label{fig:enter-label}
\end{figure}

\end{document}